\documentclass[journal]{vgtc}                     

\onlineid{1222}

\vgtccategory{Research}

\vgtcpapertype{algorithm/technique}

\title{A Task-Parallel Approach for Localized Topological Data Structures}

\author{%
  \authororcid{Guoxi Liu}{0000-0002-8164-7185},~{Student Member,~IEEE, }
  and 
  \authororcid{Federico Iuricich}{0000-0003-1782-9715},~{Member,~IEEE }
}

\authorfooter{
  \item
  	Guoxi Liu and Federico Iuricich are with School of Computing, Clemson University.
  	E-mail: \{guoxil\,$|$\,fiurici\}@clemson.edu\,.
}

\abstract{%
  Unstructured meshes are characterized by data points irregularly distributed in the Euclidian space. Due to the irregular nature of these data, computing connectivity information between the mesh elements requires much more time and memory than on uniformly distributed data. To lower storage costs, dynamic data structures have been proposed. These data structures compute connectivity information on the fly and discard them when no longer needed. However, on-the-fly computation slows down algorithms and results in a negative impact on the time performance. 
  To address this issue, we propose a new task-parallel approach to proactively compute mesh connectivity. Unlike previous approaches implementing data-parallel models, where all threads run the same type of instructions, our task-parallel approach allows threads to run different functions. Specifically, some threads run the algorithm of choice while other threads compute connectivity information before they are actually needed. The approach was implemented in the new Accelerated Clustered TOPOlogical (\textit{ACTOPO}) data structure, which can support any processing algorithm requiring mesh connectivity information. Our experiments show that ACTOPO combines the benefits of state-of-the-art memory-efficient (TTK CompactTriangulation) and time-efficient (TTK ExplicitTriangulation) topological data structures. It occupies a similar amount of memory as TTK CompactTriangulation while providing up to 5x speedup. Moreover, it achieves comparable time performance as TTK ExplicitTriangulation while using only half of the memory space.

}

\keywords{Data structures, parallel computation, topological data analysis, simplicial complex}

\graphicspath{{figs/}{figures/}{pictures/}{images/}{./}} 

\usepackage{mathptmx}                  

\usepackage{amsmath}
\usepackage{array}
\usepackage{booktabs}                   
\usepackage{caption}
\usepackage{cite}
\usepackage{float}
\usepackage{graphicx}
\usepackage{hyphenat}
\usepackage{textcomp}                   

\usepackage[svgnames]{xcolor}           

\begin{document}



\firstsection{Introduction}
\label{sec:introduction}

\maketitle

The proliferation of high-resolution scanning devices is increasing the availability and size of unstructured meshes in applications such as computer graphics \cite{Sahistan2021raytraced,Morrical2019efficient,Dai2019scan2mesh}, material science \cite{Bao2022Application,Zhao2020arbitrary}, medical modeling \cite{Shulga2017tensor,Sathar2015Tissue}, environmental science \cite{Xu2023Topology,Codd2021inversion,Heinecke2014petascale}, and autonomous navigation \cite{Hu2021spatial,Azpurua2021autonomous}.

Despite their widespread adoption, processing and visualizing unstructured meshes still represents a major bottleneck in the analysis pipeline. With regular data, computing and storing the connectivity of the mesh elements has a negligible cost since all information is implicitly provided by the data regularity. With unstructured data, instead, the same operation increases the memory footprint to the point of saturating the available memory. 

Dynamic data structures have been proposed \cite{Fellegara2021stellar,Liu2021topocluster} to cope with this problem by managing memory usage at runtime. The key idea of these approaches is to compute connectivity information only for a subset of the mesh at a time, discarding information when no longer needed. This approach provided advantages for memory consumption, but resulted in algorithms two to four times slower than state-of-the-art data structures \cite{Liu2021topocluster}. 

In this paper, we propose a new block-based task-parallel computation model to obtain data structures that are both time and memory efficient. {\em Block-based} \cite{Weiss2011prstar,Fellegara2021stellar,Liu2021topocluster} indicates a data structure that executes fine-grain operations locally (on subsets of the input mesh) rather than processing the entire mesh at once. {\em Task-parallel} \cite{Kessler2007parallel} indicates a data structure integrating pipelined data computation and data consumption tasks. The combination of these two characteristics allows the data structure to self-organize resources at runtime, computing information before they are needed for local data consumption and discarding information when no longer needed.

The main contributions of this work include:
\begin{itemize}
    \item A new task-parallel computation model for unstructured mesh processing;
    \item A new data structure implementing the proposed model;
    \item A comparison with state-of-the-art topological data structures on a wide range of unstructured meshes and processing algorithms (both sequential and parallel);
    \item An open-source integration of the proposed data structure in the TTK framework \cite{Tierny2018ttk}. 
\end{itemize}


\section{Background}
\label{sec:background}

In this section, we introduce the necessary background information for topological data structures, including the notion of simplicial complex and topological relation. 

\subsection{Simplicial complex}
A $k$-simplex (or simplex of dimension $k$) is defined as the convex hull of $k+1$ linearly independent points in the Euclidean space. A $0$-simplex is also referred to as a point, a $1$-simplex as an edge, a $2$-simplex as a triangle, and so on. Given a $k$-simplex $\sigma$, the convex hull of a nonempty subset of size $m+1$ of the $k+1$ points (i.e., $m < k$) that defines an $m$-simplex $\tau$ is called an {\em $m$-face} of $\sigma$, and $\sigma$ is said to be a coface of $\tau$. The set of cofaces of a simplex $\sigma$ forms the {\em star} of $\sigma$. 

A simplicial complex $\Sigma$ is a set of simplices such that every face of a simplex $\sigma$ is also in $\Sigma$, and the intersection of any two simplices $\sigma$ and $\tau$ is either a face of both or empty. A simplex that is not a proper face of any other simplex in $\Sigma$ is called {\em top simplex}. The {\em dimension} $d$ of $\Sigma$ is equal to the largest dimension of any simplex in $\Sigma$. 

\subsection{Topological relations}

Three types of topological relations describe the connectivity of the simplices in a simplicial complex $\Sigma$. The {\em boundary} relation maps a simplex to its faces, the {\em coboundary} relation maps a simplex to its cofaces, and the {\em adjacency} relation maps a simplex to other simplexes next to it. Suppose that two simplices $\sigma$ and $\tau$ are in $\Sigma$, and $\sigma$ is a face of $\tau$, we say that $\sigma$ is on the {\em boundary} of $\tau$, and similarly, $\tau$ is on the {\em coboundary} of $\sigma$. Two $k$-simplices $\tau_1$ and $\tau_2$ are {\em adjacent} if and only if they share a common $(k-1)$-simplex $\sigma$, and two vertices are adjacent if they are on the same edge. 

In this paper, we focus on the topological relations between the simplices of a tetrahedral mesh and use capital letters to indicate whether the relation involves a vertex ($V$), edge ($E$), triangle ($F$), or tetrahedron ($T$). Each topological relation is represented with a pair of letters, e.g., $FE$ relation denotes the edges on the boundary of a triangle. For a tetrahedral mesh, there are six boundary relations ($EV$, $FV$, $TV$, $FE$, $TE$, $TF$), six coboundary relations ($VE$, $VF$, $VT$, $EF$, $ET$, $FT$), and four adjacency relations ($VV$, $EE$, $FF$, $TT$). 

\begin{figure}[htb!]
    \centering
    \begin{tabular}{ccc}
        \includegraphics[width=0.3\linewidth]{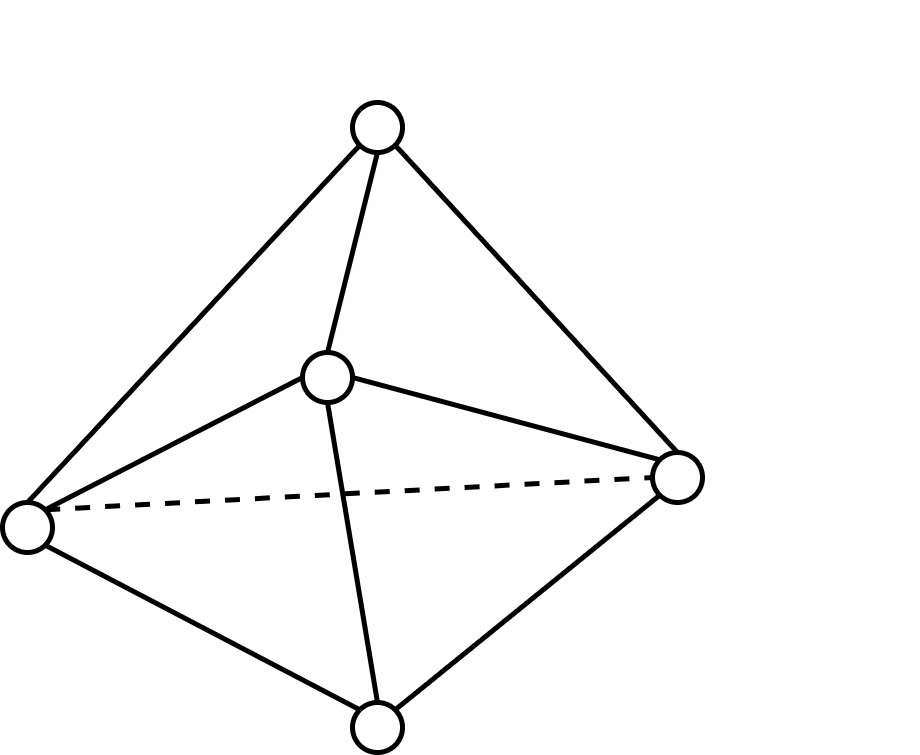} & 
        \includegraphics[width=0.3\linewidth]{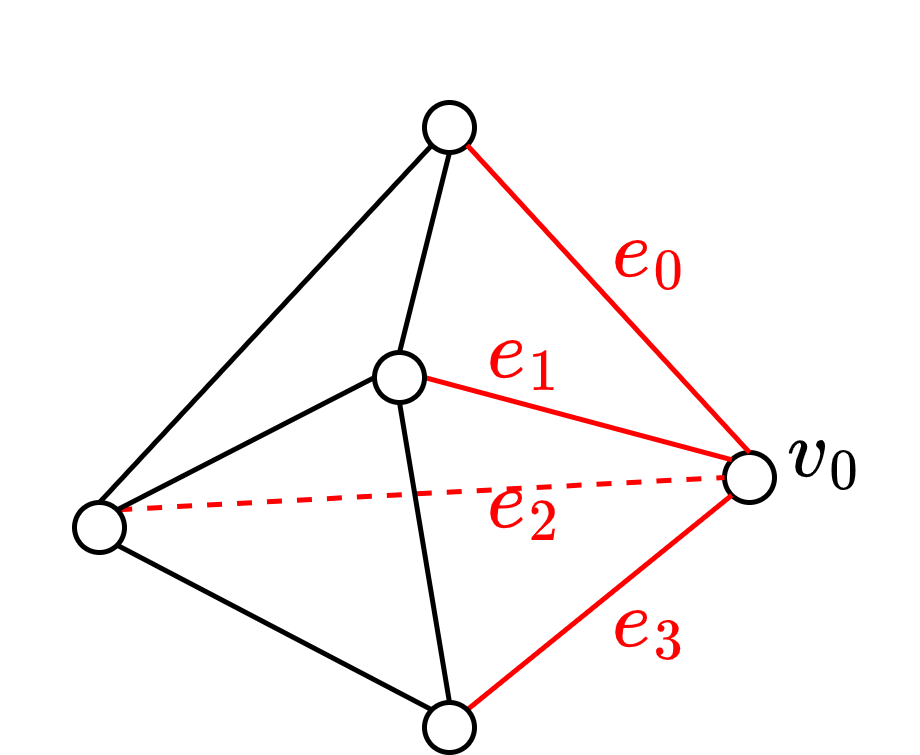} & 
        \includegraphics[width=0.3\linewidth]{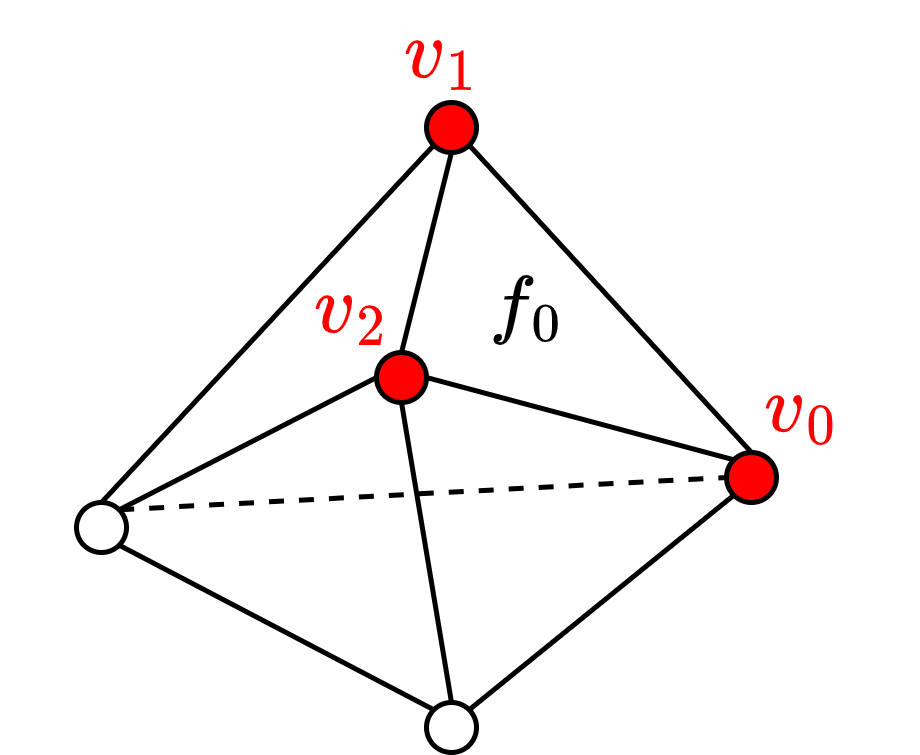} \\
         & (a) & (b) \\
    \end{tabular}
    \caption{A simplicial complex composed by two tetrahedra sharing a triangle face. (a) $VE$ relation for vertex $v_0$. (b) $FV$ relation for triangle $f_0$.}
    \label{fig:relation_example}
\end{figure}

\autoref{fig:relation_example} shows two example topological relations for a simple tetrahedral mesh composed by two tetrahedra sharing a common triangle face. \autoref{fig:relation_example}(a) shows the $VE$ relation for the vertex $v_0$, which involves the edges $e_0$, $e_1$, $e_2$, and $e_3$ highlighted in red. \autoref{fig:relation_example}(b) shows the $FV$ relation for the triangle $f_0$, which involves the vertices $v_0$, $v_1$, and $v_2$.

\section{Related work}
\label{sec:related_work}

In this section, we review data structures designed for encoding simplicial complexes, and we provide an overview of how parallel computation is used for stuying the mesh topology. 

\subsection{Data structures for simplicial complexes} 

In this work we focus on data structures that allow the retrieval of any topological relation in a mesh. A number of data structures encode unstructured meshes without representing their connectivity information. These are useful for specific tasks (e.g., ray tracing \cite{Wald2022memory}) but are inadequate to support topological algorithms. 

Data structures that provide access to topological relations can be classified into two categories: {\em static} and {\em dynamic}.

\paragraph*{Static data structures.} 
The approach adopted by static data structures is to compute and store topological relations at initialization time. Differences among them are to be found in the types of relations they encode.

The {\em Incidence graph} \cite{Edelsbrunner1987Algorithms} is the most general static data structure for simplicial complexes of arbitrary dimension, which explicitly encodes all simplices and all boundary and coboundary relations. Given the huge memory consumption it requires, several compact alternatives have been developed to reduce its memory footprint \cite{DeFloriani2004data, DeFloriani2010dimension}.  

The {\em Simplex tree} \cite{Boissonnat2014simplex} avoids encoding boundary relations by organizing all simplices of $\Sigma$ in a trie\cite{Bentley1997fast}. The result is a data structure that can efficiently support the query of coboundary relations, but that still has limited scalability when working with simplicial complexes in high dimensions \cite{Fellegara2021stellar}.

The {\em Half-edge} data structure \cite{Nielson1997tools} is a well-known data structure for triangle meshes, which reduces the storage costs by only encoding the topological relations involving edges. {\em Half-faces} \cite{Kremer2013versatile} generalizes the concept of the half-edge to polyhedral complexes. 

{\em Indexed data structures} \cite{Lawson1977software} provide a more compact option by encoding only vertices, top simplices, and the boundary relation from top simplices to their vertices. It contains sufficient information to extract efficiently all the boundary relations of cells, but it requires additional steps for coboundary or adjacency relations.

Several data structures have been developed to encode the connectivity through adjacency relations. Examples include the {\em Indexed data structure with Adjacencies (IA data structure)} \cite{Paoluzzi1993dimension, Robins2011theory} and the {\em Corner-Table} data structure \cite{Samet2006foundations} along with its several extensions specifically proposed for triangle meshes \cite{Gurung2011squad,Luffel2014Grouper} and tetrahedral meshes \cite{Gurung2010sot}. The {\em Generalized Indexed data structure with Adjacencies (IA* data structure)} \cite{Canino2011ia} extends the IA data structure to non-manifold simplicial complexes of arbitrary dimension. The IA* data structure has shown to be most compact among static topological data structures, especially as the dimension increases \cite{Canino2014representing}.

\paragraph*{Dynamic data structures.}
Unlike static data structures, dynamic data structures compute (and discard) topological relations during runtime rather than at initialization time.

The {\em PR-star octree} \cite{Weiss2011prstar} is considered the first dynamic topological data structure. It supports the reconstruction of the connectivity information of a simplicial complex by only encoding the list of tetrahedra incident in each vertex. The data structure is capable of extracting the boundary and coboundary relations locally to a subset of the mesh by using a PR-Octree decomposition of the mesh vertices. 

The {\em Stellar tree} data structure \cite{Fellegara2021stellar} generalizes the PR-star octree to handle a broader class of complexes in arbitrary dimensions and is the first concrete realization of the {\em Stellar decomposition} model \cite{Fellegara2021stellar}. The Stellar tree is shown to be more compact than most state-of-the-art static data structures, requiring only a fraction of the memory space of the latter \cite{Fellegara2021stellar}.

The Stellar decomposition model has also been adopted by the {\em TopoCluster} data structure \cite{Liu2021topocluster}, which enriches the Stellar decomposition with an implicit enumeration scheme for the mesh simplices. This scheme provides an interface for the easy integration of {\em TopoCluster} into any algorithm for topological data analysis. The easy and general integration of TopoCluster was demonstrated by deploying the data structure in the TTK framework \cite{Tierny2018ttk}, which allowed running any algorithm implemented in the framework out-of-the-box while drastically reducing the memory footprint.

\subsection{Parallel computation for topological data analysis} 

Rather than focusing on the speedup provided by the underlying data structure, some research work aimed at improving the performance of specific topological algorithms. To this end, parallel computation plays a major role in topology-based visualization \cite{DeFloriani2015Morse,Heine2016Survey,Yan2021Scalar}. 
While certain routines are embarrassingly parallel by nature, (e.g., critical points \cite{Banchoff1970critical} or Forman gradient computation \cite{Robins2011theory}), the extraction of many topological abstractions requires the study of dedicated parallel approaches. Notice that all the following methods have been evaluated on regular grids where the dataset subdivision and topological information are implicitly encoded. 

\paragraph*{Merge and contour trees.}
The {\em contour forests} algorithm \cite{Gueunet2016contour} presents a fast, shared memory multi-threaded computation of contour trees on tetrahedral meshes. The approach partitions the domain first, computes the local contour trees for each partition, and stitches the resulting forest into the final augmented contour tree. 
Gueunet et al. proposed a new approach based on Fibonacci heaps \cite{Gueunet2019task} that skips the domain subdivision step by distributing the computations of the merge tree arcs to independent tasks on the CPU cores. 
A pure data-parallel algorithm with the support of GPU acceleration, {\em Parallel Peak Pruning (PPP)} \cite{Carr2016parallel, Carr2021scalable}, has been developed for computing both merge and contour trees in unaugmented form using OpenMP for threading and using Thrust for GPU. The PPP algorithm presents up to 70x speedup compared to the serial sweep and merge algorithm supporting the contour tree computation for arbitrary (topology) graphs \cite{Carr2022distributed}. 
In contrast to building the global merge tree, the {\em merge forest} approach \cite{Klacansky2020toward} decomposes the domain, maintains the local merge trees connected by a local {\em reduced bridge set}, and computes the necessary global information only at query time. As a result, the merge forest represents a localized data structure designed for answering queries related to merge trees without expensive precomputation costs. This idea has been generalized further for distributed environments \cite{Huang2021distributed}.

\paragraph*{Morse-Smale complex.} 
Parallel algorithms for computing a 3D Morse-Smale (MS) complex \cite{Peterka2011scalable, Gyulassy2012parallel} extend the divide-and-conquer strategy presented by Gyulassy et al. \cite{Gyulassy2008practical}. The idea is to partition data into blocks, compute the MS complex for the individual blocks, and then merge the MS cells with a dedicated merge-and-simplify routine. Many approaches have focused on the geometric accuracy of the reconstructed model rather than the efficiency of the parallel approach \cite{Bhatia2018topoms, Gyulassy2012computing, Gyulassy2019shared, Gyulassy2014conforming}. 
An exception is the hybrid (CPU-GPU) shared-memory algorithm proposed by Shivashankar et al. \cite{Shivashankar2012parallel}. The algorithm assigns embarrassingly parallel tasks such as gradient computation and extreme traversals to the GPU, and thus results in substantial speedup over CPU-based approaches. 
A pure GPU parallel algorithm for computing the MS complex has also been developed recently \cite{Subhash2020gpu}, which transforms the graph traversal operations into vector and matrix operations that are better suited for GPU parallel computation. 

\section{Task-parallel computation approach}
\label{sec:methods}

In this section, we describe our proposed task-parallel model. To clarify the difference between our proposed model and existing approaches, we first review common computation models used by static and dynamic data structures.

\begin{figure}[htb!]
    \centering
    \includegraphics[width=\linewidth]{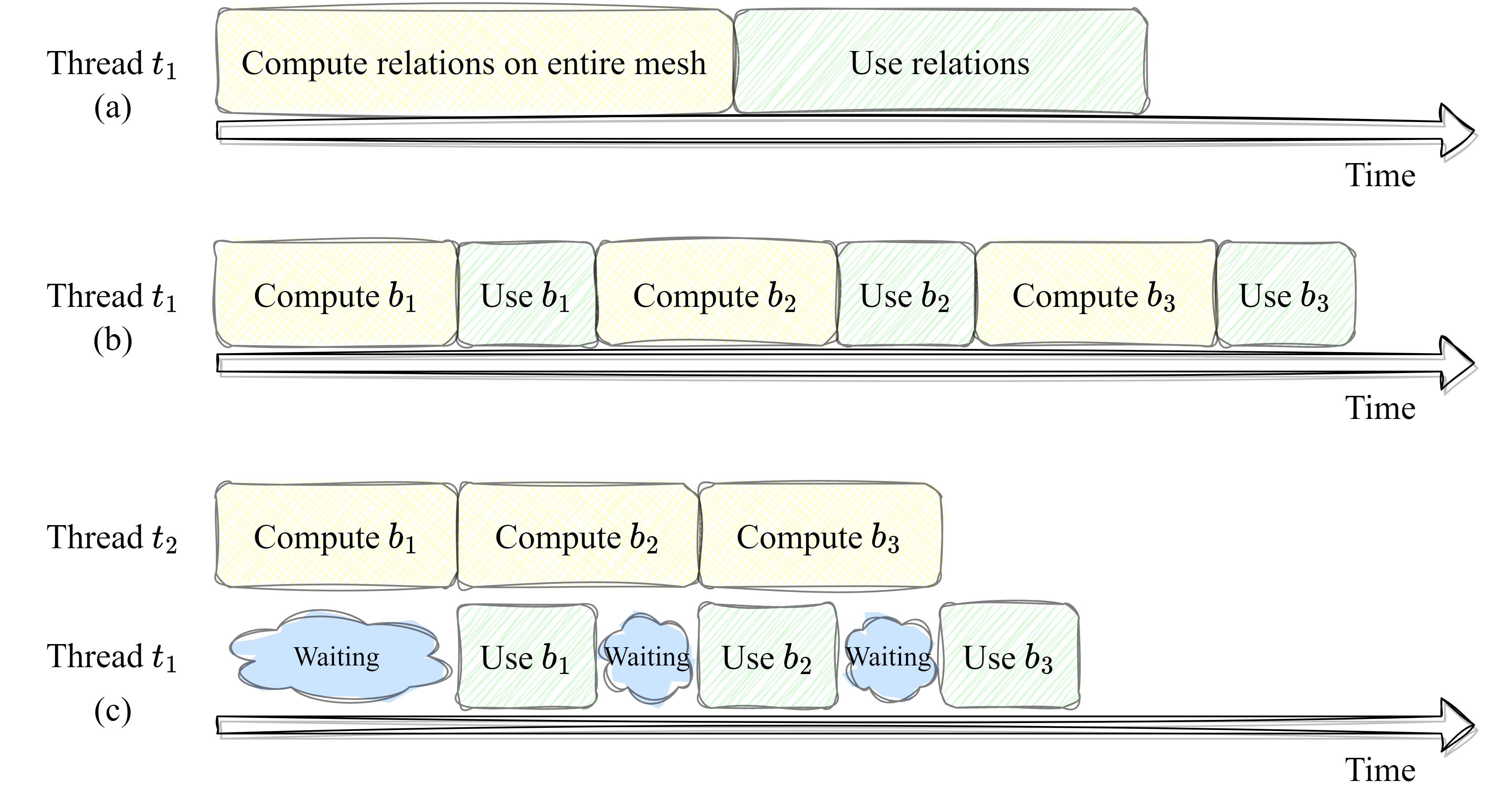}
    \caption{(a) Workflow of typical static data structures. The thread computes topological relations for the entire mesh first and then use them in the algorithm. (b) Workflow of the traditional dynamic approach. The thread works with one block at a time. It computes and uses the topological relations in a sequential order. (c) Workflow of the proposed task-parallel approach with two threads. One is only responsible for precomputing topological relations, and the other executes the algorithm by using the topological relations.}
    \label{fig:static_dynamic_methods}
\end{figure}

Static data structures use a preprocessing approach, where topological relations are computed in bulk during the initialization of the data structure and then stored until the algorithm has been executed. As shown in \cref{fig:static_dynamic_methods}(a), thread $t_1$ first computes the topological relations of the entire mesh, and then uses them in the selected algorithm. 

Dynamic data structures compute information on-the-fly by processing one subset of the mesh at a time, which is achieved by dividing the dataset into blocks \cite{Weiss2011prstar, Fellegara2021stellar, Liu2021topocluster}. In practice, the running thread accesses one of the blocks, computes the required topological relations, and then runs the algorithm inside the block. \cref{fig:static_dynamic_methods}(b) shows an example of such workflow executed by thread $t_1$ that computes the topological relations for the mesh subset $b_1$, executes the algorithm locally on $b_1$ using these relations, and then moves to the next portion of the mesh $b_2$ to repeat the same process. 

The goal of our task-parallel model is to avoid the switching between algorithm execution and topological relations computation. Our model still assumes that the input mesh is subdivided into blocks. However, it introduces two types of threads, specialized in the computation of topological relations and in the algorithm execution, respectively. \cref{fig:static_dynamic_methods}(c) shows an example of our proposed approach. Thread $t_1$ is created for running the processing algorithm, while thread $t_2$ is used for computing topological relations. In practice, thread $t_2$ precomputes topological relations for thread $t_1$ so that, as long as topological relations are provided, $t_1$ keeps executing instructions from the selected algorithm on each block.

The key difference in the execution of thread $t_1$ is that the time originally spent to compute topological relations in the traditional approach (\cref{fig:static_dynamic_methods}(b)) is now replaced by a waiting time (\cref{fig:static_dynamic_methods}(c)) in the new approach. The more efficient thread $t_2$ computes topological relations for $t_1$, the shorter $t_1$'s waiting time will be.

\subsection{Consumers and producers}
\label{sec:consumer_producer}

Our approach integrates a classic producer-consumer paradigm with constrained consumers \cite{Hilzer1992Synchronization}. Threads that compute topological relations are called {\em producers}. Threads that use these topological relations to run a processing algorithm are called {\em consumers}.

Our model involves two types of producers. {\em Worker producers} are unconstrained threads that can compute topological relations in any block without a specific request. {\em Leader producers} are constrained producers that compute topological relations only if specifically asked by a consumer. Moreover, they manage the communication between consumers and worker producers.

\begin{figure}[htb!]
    \centering
    \includegraphics[width=\linewidth]{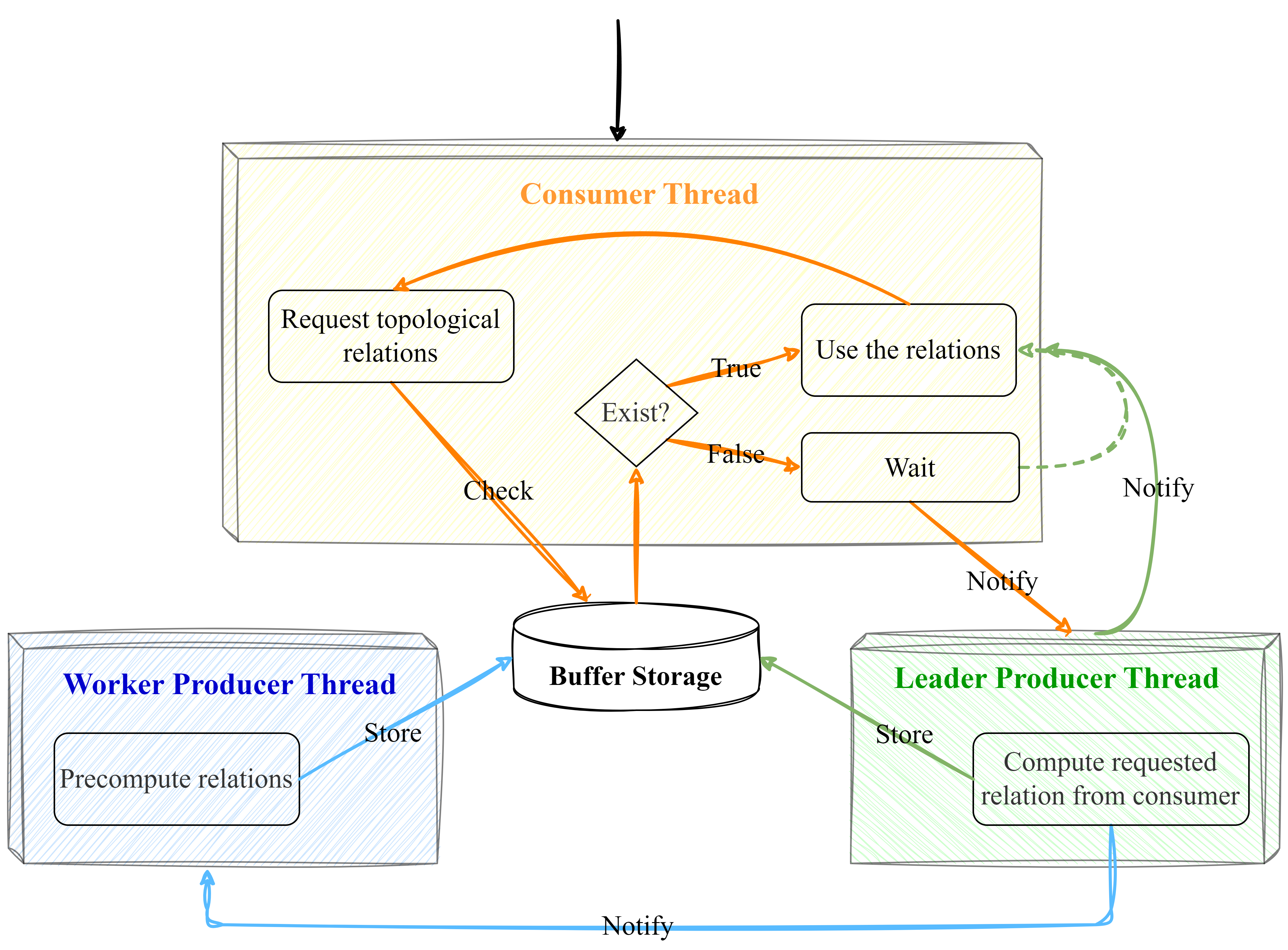}
    \caption{The task-parallel producer-consumer paradigm. Orange, green, and blue colors indicate the activities of the consumer, leader producer, and worker producer threads, respectively.}
    \label{fig:pipeline}
\end{figure}

\cref{fig:pipeline} shows the general workflow and communication strategy used by these types of threads. In the following, we provide a detailed description of the behavior of each thread type, assuming a sequential algorithm is being executed (i.e., a single consumer thread is involved).

\paragraph*{Consumer.}
The consumer thread is the first thread to be spawned. The consumer thread spawns the leader producer thread and runs the selected processing algorithm that makes requests for topological relations. 
Our model involves a buffer system that all threads can use to store/retrieve topological relations (see \cref{fig:pipeline}). Assuming that a topological relation $R$ for a simplex $\sigma$ is requested, the consumer first identifies the block $b_i$ containing $\sigma$. Then, it searches the buffer to verify if the relation $R$ for the block $b_i$ has been computed. If yes, the consumer keeps running the processing algorithm inside $b$. Otherwise, the consumer stops its execution and invokes a request to the leader producer. 

\paragraph*{Leader producer.}
The leader producer acts as a middle-layer between the consumer and the worker producers. During the initialization phase, the leader producer creates the worker producers and starts monitoring the consumer's working block (i.e., on which block the consumer is running the processing algorithm). 

Every time the consumer requests a topological relation for a simplex $(e.g., R(\sigma))$ that is not stored in the buffer, the leader producer promptly computes the relation $R$ for the block $b_i$ containing the simplex $\sigma$ and notifies the consumer that the topological relation $R(\sigma)$ is ready. Then, the leader producer notifies the worker producers that topological relations will be needed in the proximity of $b_i$. 

\paragraph{Worker producer.}
Similar to the leader producer, the task of worker producers is to precompute topological relations in a block and store them in the buffer. The difference is that worker producers process blocks independently without an explicit request, unless they are notified by the leader producer.

Worker producers are spawned by the leader producer and are allocated to a block $b_i$ to compute topological relations. Worker and leader producers share a variable indicating the index $i$ of the block $b_i$ where the consumer is working. Moreover, they share a second variable indicating the index $j$ of the last topological relation $R_j$ requested by the consumer. These variables are controlled by a lock to prevent multiple producers from computing the same topological relation on the same block. Once a worker thread acquires the lock to access $b_i$, it modifies either $b_i$ or $R_j$ to the new value for next worker, releases the lock, and then proceeds with computing the topological relation with saved $R_j$ and $b_i$ before modification. This way, the next worker thread that acquires the lock will not compute the same relation on the same block (see Section \ref{sec:worker_mode} for details on how worker producers update the block $b_i$ and topological relation index $R_j$).

This workflow can only get interrupted by the leader producer at any time, which will force all worker threads to move to $b_{i'}$ by updating the corresponding shared variable.

\subsubsection{Computing new blocks for worker producers}
\label{sec:worker_mode}

After studying a wide range of processing algorithms, particularly those related to topology-based visualization, we have recognized key differences in their access patterns (the order in which they access blocks) and the types of topological relations they request. To address this diversity, we have defined multiple computing modes for worker producers to experiment with different orders for visiting blocks and the number of topological relations to compute. All computing modes have been implemented and experimentally evaluated (see Section \ref{sec:evaluation_single}).

\paragraph*{Moving directions.}
As discussed previously, worker producers control the exclusive access to blocks through a shared variable. We have defined two possible strategies that a worker thread can use to update the variable.

\begin{figure}[bthp]
    \centering
    \begin{tabular}{cc}
        \includegraphics[width=0.4\linewidth]{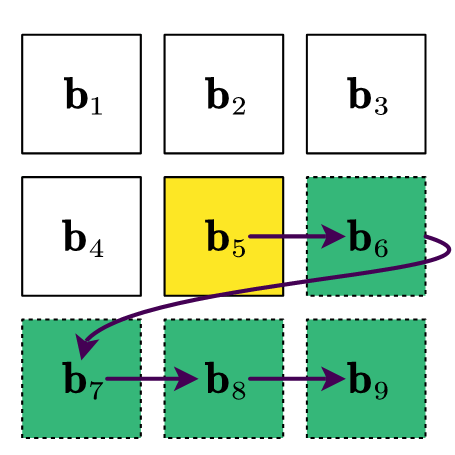} & 
        \includegraphics[width=0.4\linewidth]{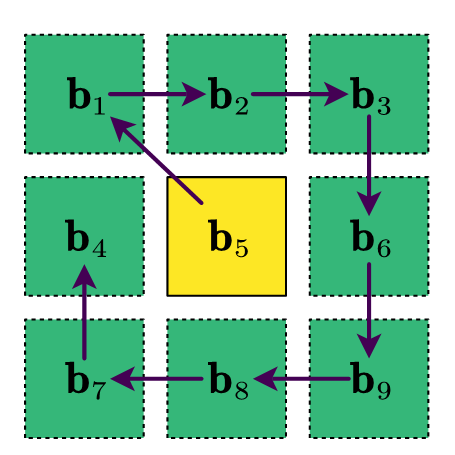} \\
        (a) & (b) \\ 
    \end{tabular}
    \caption{Different moving directions (in green color) of worker producers for block $b_5$: (a) linear order, (b) spatial order.}
    \label{fig:linear_spatial_order}
\end{figure}

The first strategy aims at precomputing relations for blocks with the index following the block $b_i$, i.e., the next block to compute is $Next(b_{i}) = b_{i+1}$. The strategy is designed based on the fact that many topology-based visualization algorithms loop through the simplices in the mesh based on their indices. In most block-based data structures \cite{Weiss2011prstar,Fellegara2021stellar,Liu2021topocluster}, block indices follow the same order as the global indices of simplices.
We refer to this buffer strategy as {\em linear} buffer, which is demonstrated in \cref{fig:linear_spatial_order}(a). Given the current block is $b_5$, the worker thread will precompute block $b_6$, followed by $b_7$, $b_8$, and $b_9$. 

The second strategy aims at precomputing relations for all neighbors of a block $b_i$. We say two blocks are neighbors if they share a common tetrahedron. This strategy is motivated by the fact that certain algorithms (e.g. Morse-Smale computation \cite{Edelsbrunner2003morse}) visit simplices based on their connectivity rather than their indexing. These algorithms visit neighboring blocks in an unpredictable order which motivates the precomputation of all blocks surrounding the current one. We refer to this buffer strategy as {\em spatial} buffer. \cref{fig:linear_spatial_order}(b) shows an example of the moving direction in the spatial order. Since the neighbor set of the current block $b_5$ is $Neighbors(b_5) = \{b_1, b_2, b_3, b_4, b_6, b_7, b_{8}, b_{9}\}$, after $b_5$ is computed, the worker thread will start visiting blocks in $Neighbors(b_5)$, e.g., $b_1$. 

\paragraph*{Topological relations.}
After moving to a new block $b_i$, the worker producer will compute topological relations $R$ for the simplices in $b_i$. At this point, the worker thread has two options, and the most straightforward one is to compute only the topological relation $R$ specified in the last consumer's request. In this case, the worker producer will update the block index once finished (based on the moving direction mentioned earlier) while leaving the topological relation index $R_j$ untouched.

However, we observed that processing algorithms rarely use only one topological relation at a time. Instead, multiple topological relations are usually requested for the same block or for the same simplex. Therefore, an alternative approach for the worker producer is to compute all types of topological relations used by the algorithm when accessing a new block.
In this case, each worker producer will update the relation index $R_j$ to indicate that more topological relations need to be extracted in this block. Once all topological relations are precomputed, the last worker thread accessing the block will increment the block index $b_i$, according to the specified moving direction.

\section{The ACTOPO Data Structure}
\label{sec:seq_algorithm}

Using the task-parallel framework described in \cref{sec:methods}, we implement a new data structure called Accelerated Clustered TOPOlogical ({\em ACTOPO}) data structure. In the remainder of this section, we describe the encodings of ACTOPO and the implementation choices made to integrate consumer, leader producer, and worker producer threads in the data structure.

\subsection{Encodings of static information}
\label{sec:encodings}

The static information encoded by ACTOPO comprises the representation of the input unstructured mesh and additional information computed at initialization time. In the following, we assume that the input is a tetrahedral mesh that has already been subdivided into blocks. However, the same approach could be used to encode meshes defined by general polytopes.

\paragraph*{Input.}
The input tetrahedral mesh contains only information for vertices and tetrahedra. An indexed list $V$ stores the coordinate values of each vertex. An indexed list $T$ stores the vertices forming each tetrahedron ($TV$ relation). 

The data structure also assumes a subdivision of the mesh is defined based on the mesh vertices and is also provided in the input. Vertices with the same label belong to the same {\em block}. An indexed list $I$ is used to encode the block that each vertex belongs to. 
It is a requirement of the subdivision that each vertex is contained in exactly one block.

\begin{figure}[htb!]
    \centering
    \includegraphics[width=0.8\linewidth]{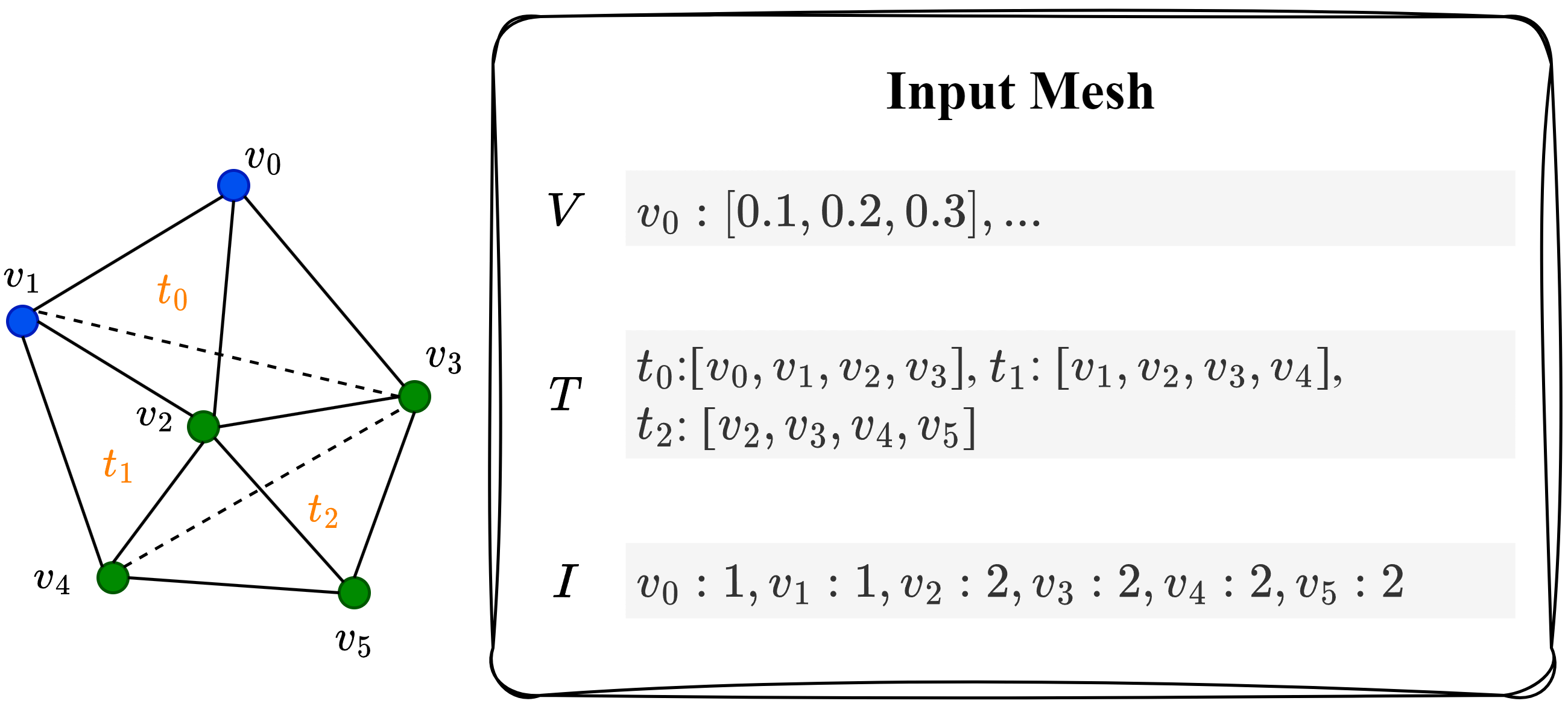}
    \caption{The input mesh that includes the vertex list $V$, tetrahedron list $T$, and a subdivision $I$.}
    \label{fig:encoding_input}
\end{figure}

\cref{fig:encoding_input} shows an example of the encodings of an input mesh. The example tetrahedral mesh shows that $v_0$ and $v_1$ belong to the same block (i.e., $I[v_0] = I[v_1] = b_1$), while the remaining vertices are in the block $b_2$ (i.e., $I[v_2] = I[v_3] = I[v_4] = I[v_5] = 2$).

Based on the subdivsion of the mesh vertices we can create an association between a simplex $\sigma$ and a block $b_i$. Specifically, we say that a simplex $\sigma$ is {\em internal} to the block $b_i$ iff. the vertex $v$ of $\sigma$ with the lowest index also belongs to $b_i$ and is {\em external} to all other intersecting blocks. 

\paragraph*{Initialization.} 
In addition to the input mesh, the data structure computes and stores the boundary relations of each simplex $\sigma$ with its vertices. Since this information is not provided for edges and triangles (i.e., $EV$, $FV$), these relations are computed during the initialization phase. 

\begin{figure}[htb!]
    \centering
    \includegraphics[width=\linewidth]{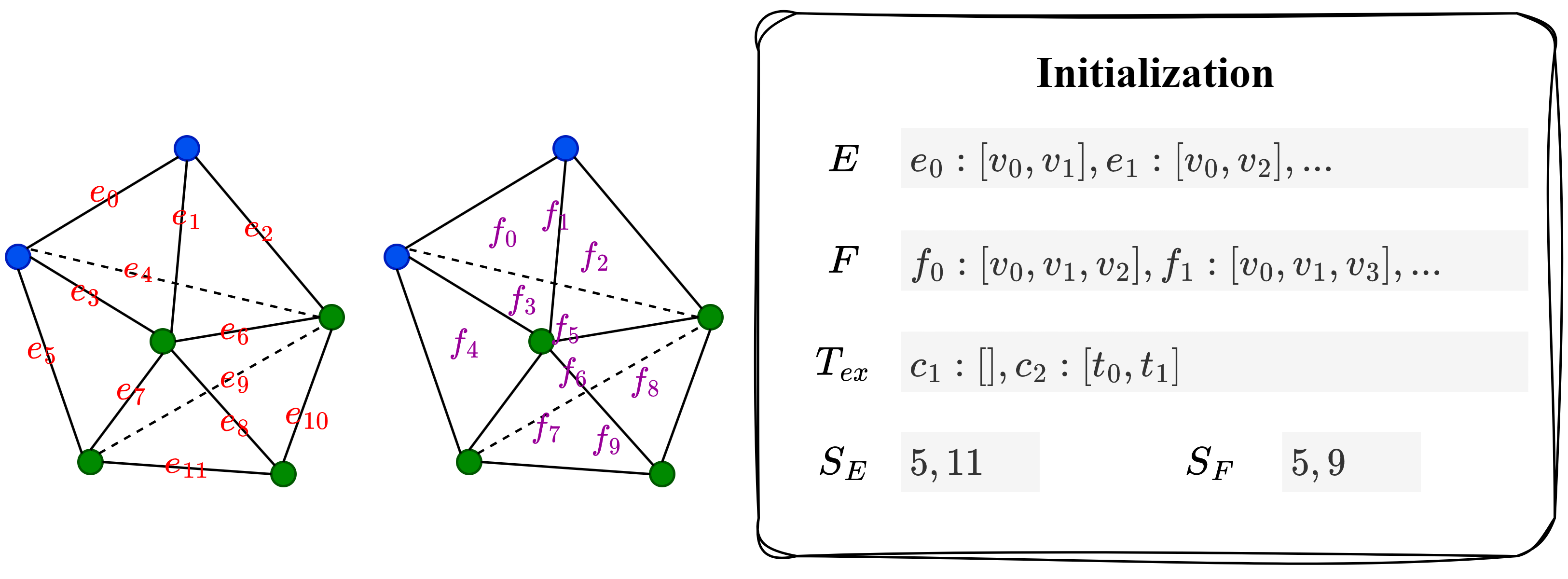}
    \caption{Encodings of ACTOPO after initialization, which include edge list $E$, triangle list $F$, external tetrahedra list $T_{ex}$, and interval arrays for edges and triangles $S_E$ and $S_F$.}
    \label{fig:encoding_initial}
\end{figure}

The extraction is performed by iterating through the tetrahedra in each block. For each tetrahedron, all possible combinations of two vertices ($EV$ relation) and three vertices ($FV$ relation) are collected and stored in an indexed list after removing duplicates. \cref{fig:encoding_initial} shows the additional information encoded by the data structure after the initialization. Two indexed lists $E$ and $F$ are used to store the edges and triangles globally. For each block $b_i$, we also encode a list of external tetrahedra $T_{ex}$.

In addition, the data structure maintains an interval array for edges and triangles, $S_E$ and $S_F$ respectively, indicating the largest edge and triangle indices contained in each block. These arrays are used to get local edges and triangles of a block from the global edge and triangle lists $E$ and $F$.

\subsection{Operations supported at runtime}

As mentioned earlier in \cref{sec:consumer_producer}, every time the algorithm requires a topological relation of a simplex $\sigma$, the consumer will first locate the block $b_i$ containing $\sigma$. This step requires constant time, as $b_i = I(v)$ for a vertex $v=\sigma$ and $b_i = min_{v \in \sigma} I(v)$ for any other simplex $\sigma$. 

The buffer system is implemented using a linked list that stores the index $i$ of each precomputed block $b_i$. While the linked list ensures fast insertion and deletion operations, a supplemental hash table of block indices is used for quick lookup. Therefore, it just takes constant time for the consumer thread to check whether a block $b_i$ is available in the buffer. From the consumer thread's perspective, the buffer works like a black box. This feature makes the model universally applicable to any topological algorithm.

While the consumer thread can only read data from the buffer, the leader producer is the thread that actually manages the buffer. The buffer system in ACTOPO operates on a First-In-First-Out (FIFO) order, and an additional integer value caps the maximum number of blocks allowed in the buffer. Before notifying worker producers, the leader producer checks the free space remaining in the buffer. If the buffer is at capacity, the leader producer cleans up half of the buffer based on the chronological order in which blocks were added.

It is worth noting that the computation of a local topological relation $R$ within a block $b_i$ is not different from the approach used by other dynamic data structures \cite{Fellegara2021stellar,Liu2021topocluster}. Typically, this involves iterating through internal, and sometimes external, tetrahedron list of $b_i$, and constructing the simplices of interest from vertices of the tetrahedron (e.g, a pair of vertices can form an edge). Due to the limited space available, we report the pseudocode used to extract a representative set of these operations in supplemental materials.

\section{Evaluation of Performance}
\label{sec:evaluation_single}

In this section, we present an experimental evaluation of our proposed data structure when used in combination with sequential algorithms. All experiments are performed on a desktop computer equipped with a 3.2 GHz Intel i7-8700 CPU and 32 gigabytes of RAM. We report the average values obtained over 10 runs for each experiment.

\subsection{Experimental setup}
The tetrahedral meshes selected for the performance analysis are listed in \cref{tab:datasets}. Datasets Shapes and Hole are irregular tetrahedral meshes. The remaining datasets (i.e., Red Sea, Engine, Earthquake, Foot, Asteroid, and Stent) are tetrahedral meshes created by tetrahedralizing volume images while filtering out elements corresponding to empty parts of the original data domain.

\begin{table}[!htb]
    \centering
    \caption{Overview of the experimental datasets, including the number of vertices $|V|$, edges $|E|$, triangles $|F|$, and tetrahedra $|T|$. The type field indicates whether the dataset was originally a volume image (\textit{Regular}) or points are irregularly distributed within the domain (\textit{Irregular}).}
    \label{tab:datasets}
    \resizebox{\columnwidth}{!}{
        \begin{tabular}{llrrrr}
        \toprule
        Dataset & \multicolumn{1}{c}{Type} & \multicolumn{1}{c}{$|V|$} & \multicolumn{1}{c}{$|E|$} & \multicolumn{1}{c}{$|F|$} & \multicolumn{1}{c}{$|T|$} \\ \midrule
        Red sea & Regular & 0.95M & 6.33M & 10.58M & 5.20M \\ 
        Engine & Regular & 1.39M & 9.14M & 15.18M & 7.43M \\ 
        Earthquake & Regular & 1.62M & 11.14M & 18.92M & 9.41M \\
        Foot & Regular & 4.60M & 30.79M & 51.51M & 25.32M\\ 
        Asteroid & Regular & 5.07M & 35.00M & 59.58M & 29.65M \\
        Shapes & Irregular & 7.87M & 52.37M & 87.63M & 43.13M \\ 
        Hole & Irregular & 9.26M & 63.70M & 108.29M & 53.85M \\ 
        Stent & Regular & 17.37M & 118.79M & 201.40M & 99.98M \\ \bottomrule
    \end{tabular}
    }
\end{table}
    
ACTOPO has been implemented as a new module of the Topology Toolkit (TTK version 1.1.0), and thus all plugins implemented in TTK can run seamlessly using the new data structure. We used four different TTK plugins to evaluate the performance under different runtime conditions.

\textbf{\texttt{Test\-Topo\-Relations} plugin} is a plugin developed in-house for profiling the data structure under simplified conditions. The plugin iterates over the simplices of the mesh starting from the vertices and successively moving to edges, triangles, and tetrahedra. For each simplex $\sigma$ the plugin requires the computation of every boundary and coboundary relation involving $\sigma$. The plugin traverses the tetrahedral mesh following the linear order of simplex indices, i.e., from $\sigma_0$ to $\sigma_n$, where $\sigma$ denotes a $k$-simplex in the tetrahedral mesh and $n$ is the total number of $k$-simplices in the mesh. It is selected to test the performance of the topological data structure in computing topological relations one at a time and to avoid recomputation for localized topological data structures since the topological relations of a block will only be requested once.

\textbf{\texttt{Scalar\-Field\-Critical\-Points} plugin} is used to compute critical points based on an input scalar function. The plugin requires topological relations involving the vertices (i.e., $VV$, $VF$, and $VT$). The only exception is the $FT$ relation, which is used to identify the list of vertices on the boundary of the mesh. The plugin traverses the vertices of the mesh in the same linear order as \texttt{Test\-Topo\-Relations} plugin, and then marks vertices that are on the boundary of the mesh. However, the plugin requires multiple topological relations to be computed when visiting one block instead of only one relation as in the previous testing plugin. 

\textbf{\texttt{Discrete\-Gradient} plugin} \cite{Forman2001user} computes a {\em discrete Morse gradient field} based on an input function. The function $F$ is a discrete Morse function if for any $p$-simplex $\sigma$, all the $(p-1)$-simplices on its boundary have a lower $F$ value and all the $(p+1)$-simplices on its coboundary have a higher $F$ value, with at most one exception. If such exception exists, it defines a pair of cells called a {\em discrete gradient vector}. Otherwise, $p$-simplex $\sigma$ is a {\em critical} simplex of index $p$. Intuitively, a discrete vector field can be viewed as a collection of arrows, connecting a $p$-simplex of mesh $\Sigma$ to an incident $(p+1)$-simplex in such a way that each simplex is a head, or a tail of at most one arrow and the critical simplices are neither the head nor the tail of any arrow. The plugin iterates all the $0$-simplices (i.e., vertices), and for each vertex, it adds all $k$-simplices ($k > 0$) in the lower star of the vertex into a list. If the list is not empty, the plugin finds the pairable 1-simplex (i.e., edge) for the vertex, 2-simplices (i.e., triangles) for the remaining 1-simplices, and so on. Multiple topological relations are used when computing the discrete gradient vector, specifically, $VE$, $VF$, $VT$, $EF$, $ET$, $FE$, $FT$, and $TF$ relations. Even though the plugin visits vertices sequentially, a simplex can be the face/coface of multiple simplices, and the recomputation of its topological relations will be requested multiple times during the process. 

\textbf{\texttt{Morse\-Smale\-Complex} plugin} \cite{Shivashankar2012parallel} computes a Morse-Smale (MS) complex from an input scalar function on the given tetrahedral mesh. An {\em integral line} is a path on the mesh that is tangent to the gradient of the function everywhere. Intuitively, the {\em MS complex} is a segmentation of the input scalar field in regions where integral lines are connected to the same pair of critical points. 
The plugin first computes the discrete gradient on the tetrahedral mesh as described in the \texttt{Discrete\-Gradient} plugin, and the remaining steps require to visit the mesh based on the discrete gradient vector, including the computation of 1-separatrices, saddle connectors, and segmentation. The order in which simplices are visited is defined at runtime. For this reason, this plugin denotes the worst-case scenario of the localized data structure, as recomputation of one same block could happen multiple times during the process. Since the first step of the plugin is overlapped with the previous \texttt{Discrete\-Gradient} plugin, we measure its performance by only counting the steps after the discrete gradient computation. 

\subsection{Evaluating robustness on sequential algorithms}

In this section, we test how robust the proposed approach is when varying the model's parameters, namely, the four different computing modes and the total number of producers. For each execution, we track the amount of time to finish the algorithm execution and the peak memory usage observed during the process. Notice that the execution time of the plugin includes the waiting time of the consumer (i.e., time spent waiting for a topological relation to be ready) and algorithm execution time (i.e., time spent executing the algorithm's instructions). Since parameter changes in the data structure do not affect the algorithm, a reduction in the overall execution time is attributed to a reduction in the waiting time.

\subsubsection{Comparison of computing modes for worker producers}
\label{sec:results_mode}

The working modes described in \cref{sec:worker_mode} generate four possible configurations for the worker producers. Producers can visit blocks either linearly or following a spatial order. Moreover, at each request, producers can either compute the requested topological relation or all topological relations predefined by the algorithm. Our hypothesis is that each mode may be preferable based on the algorithm's characteristics. Specifically, the linear computing mode should benefit plugins that iterate through simplices in sequential order, i.e., \texttt{Test\-Topo\-Relations}, \texttt{Scalar\-Field\-Critical\-Points}, and \texttt{Discrete\-Gradient} plugins. The spatial computing mode should be beneficial for plugins that visit simplices with an unpredictable order (e.g., \texttt{Morse\-Smale\-Complex} plugin). In the following experiments, we focus on the computing mode and test these hypotheses using 6 producers in total (i.e., 1 leader producer and 5 worker producers).

\begin{figure*}[!htb]
    \centering
    \newcolumntype{A}{>{\centering\arraybackslash} m{0.3\linewidth}}
    \begin{tabular}{A A A}
        \toprule
        \texttt{Test\-Topo\-Relations} & \texttt{Scalar\-Field\-Critical\-Points} & \texttt{Discrete\-Gradient}\\
        \midrule \\ 
        \includegraphics[width=\linewidth]{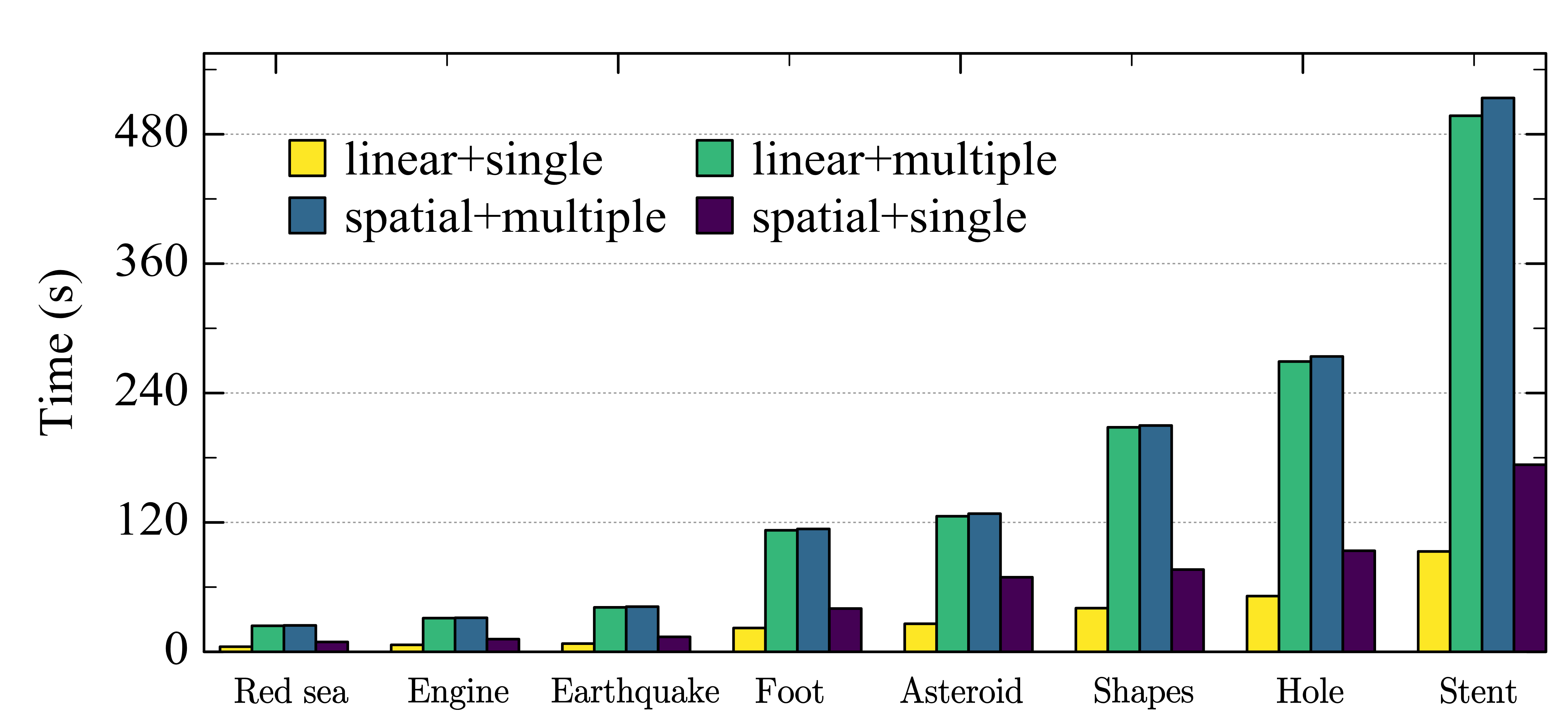} & 
        \includegraphics[width=\linewidth]{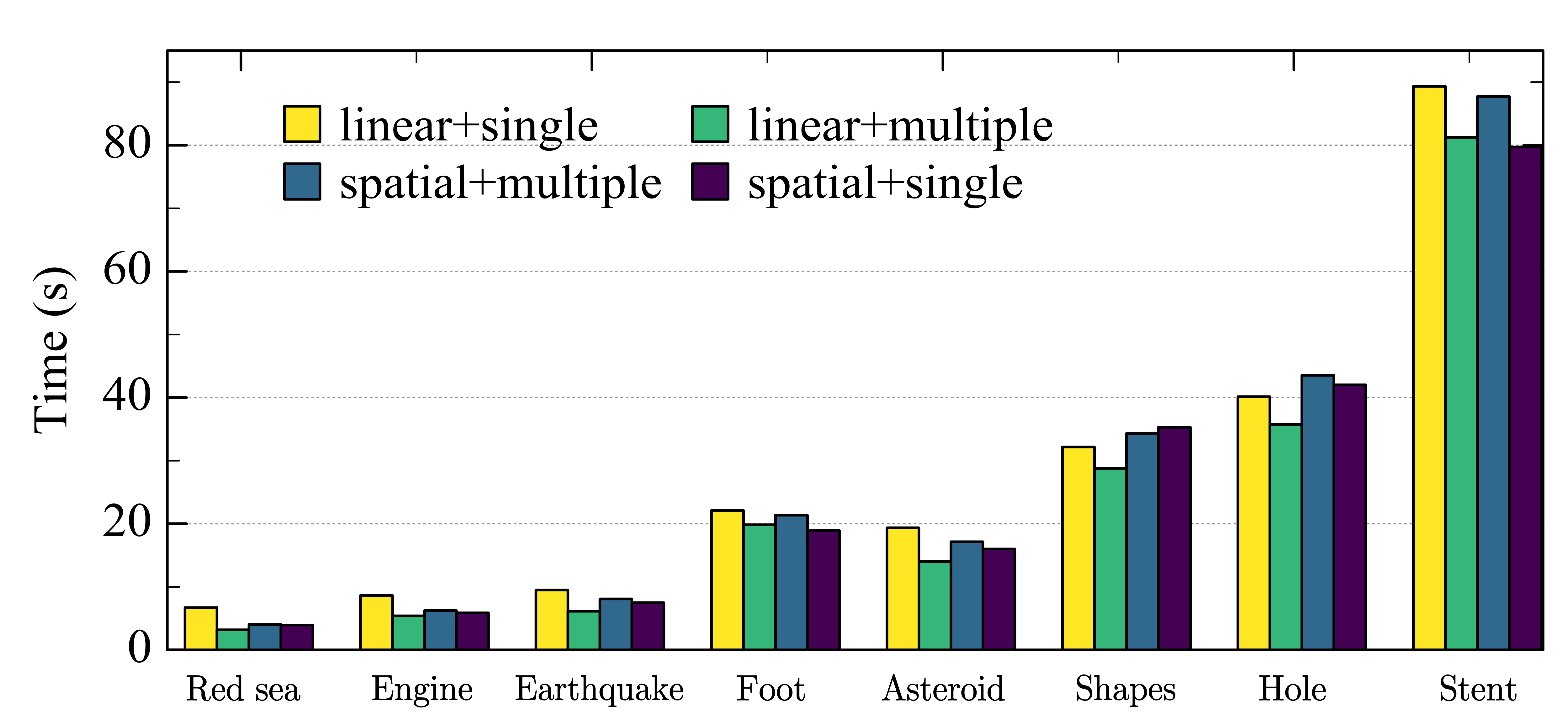} & 
        \includegraphics[width=\linewidth]{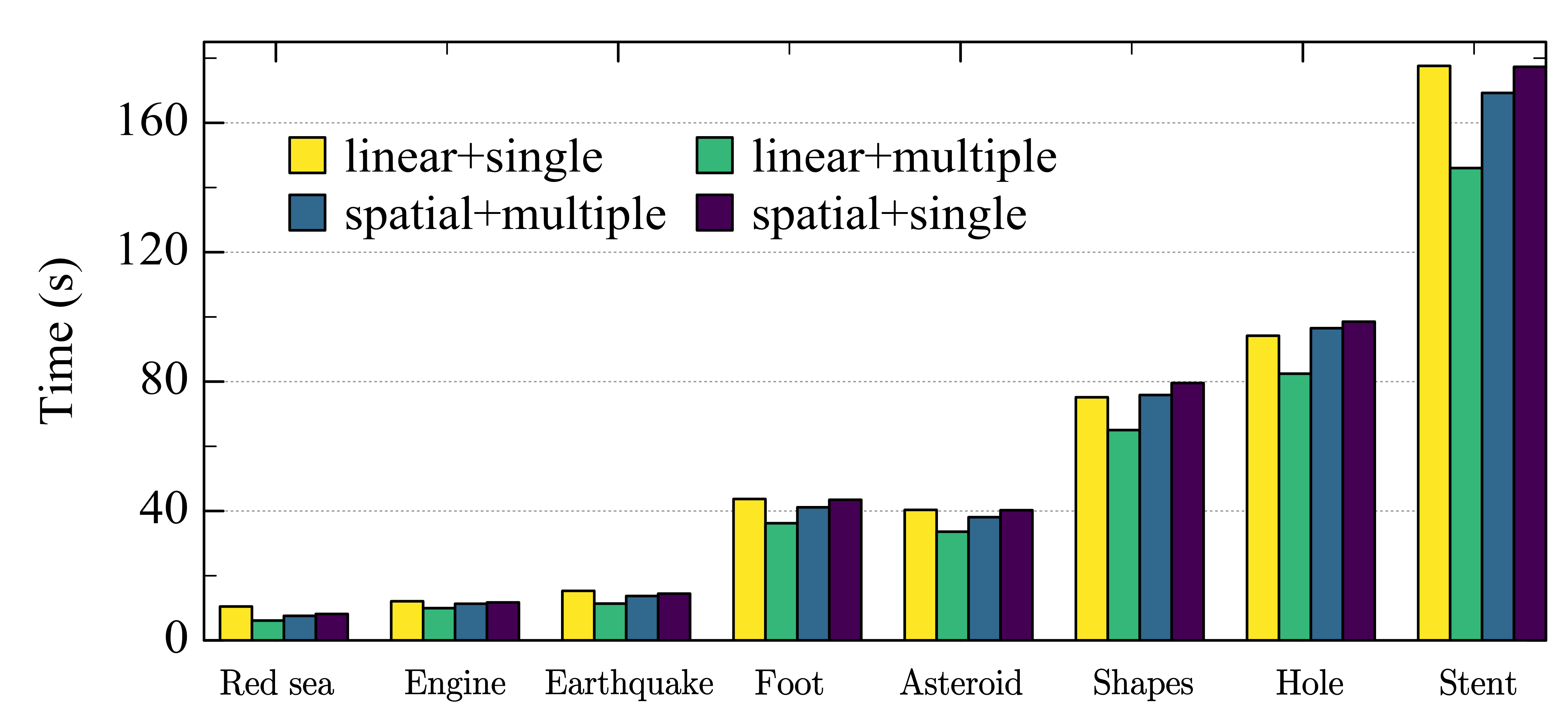} \\ 
        (a) & (b) & (c) \\
        \bottomrule
    \end{tabular}
    \caption{The execution time (in seconds) used by \texttt{Test\-Topo\-Relations}, \texttt{Scalar\-field\-Critical\-Points}, and \texttt{Discrete\-Gradient} plugins when running with different computing modes using six producer threads.}
    \label{fig:res_mode}
\end{figure*}

\cref{fig:res_mode} shows the results obtained on three plugins where the linear buffer was expected to provide the best performance. 

For \texttt{Test\-Topo\-Relations} plugin (\cref{fig:res_mode}(a)), we can notice that the topological relation computation makes the real difference. While the linear buffer provides a small benefit compared to the spatial buffer, computing only the required topological relation provides a 5.2 times speedup compared to computing all topological relations used by the algorithm.

\begin{figure}[htb!]
    \centering
    \begin{tabular}{cc}
    \includegraphics[width=0.45\linewidth]{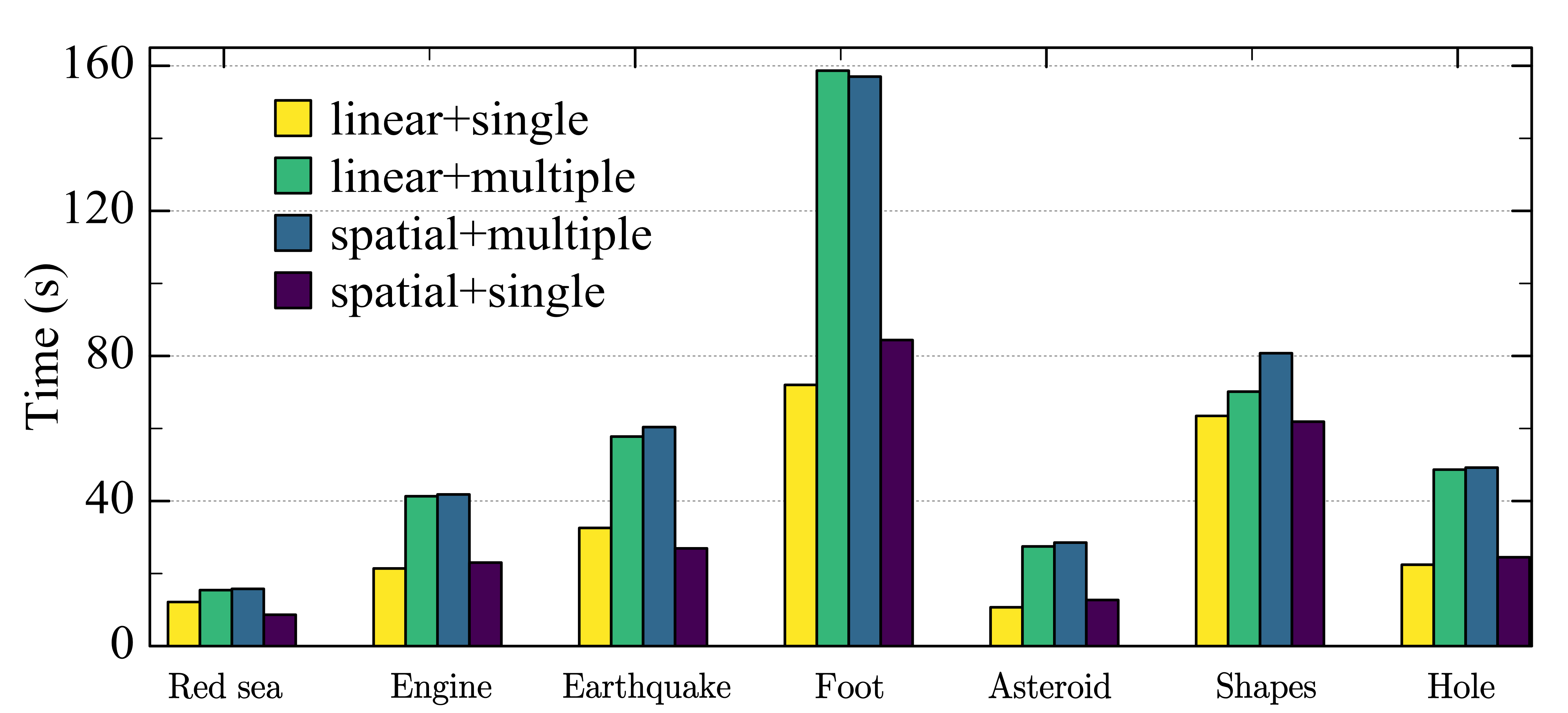} &
    \includegraphics[width=0.45\linewidth]{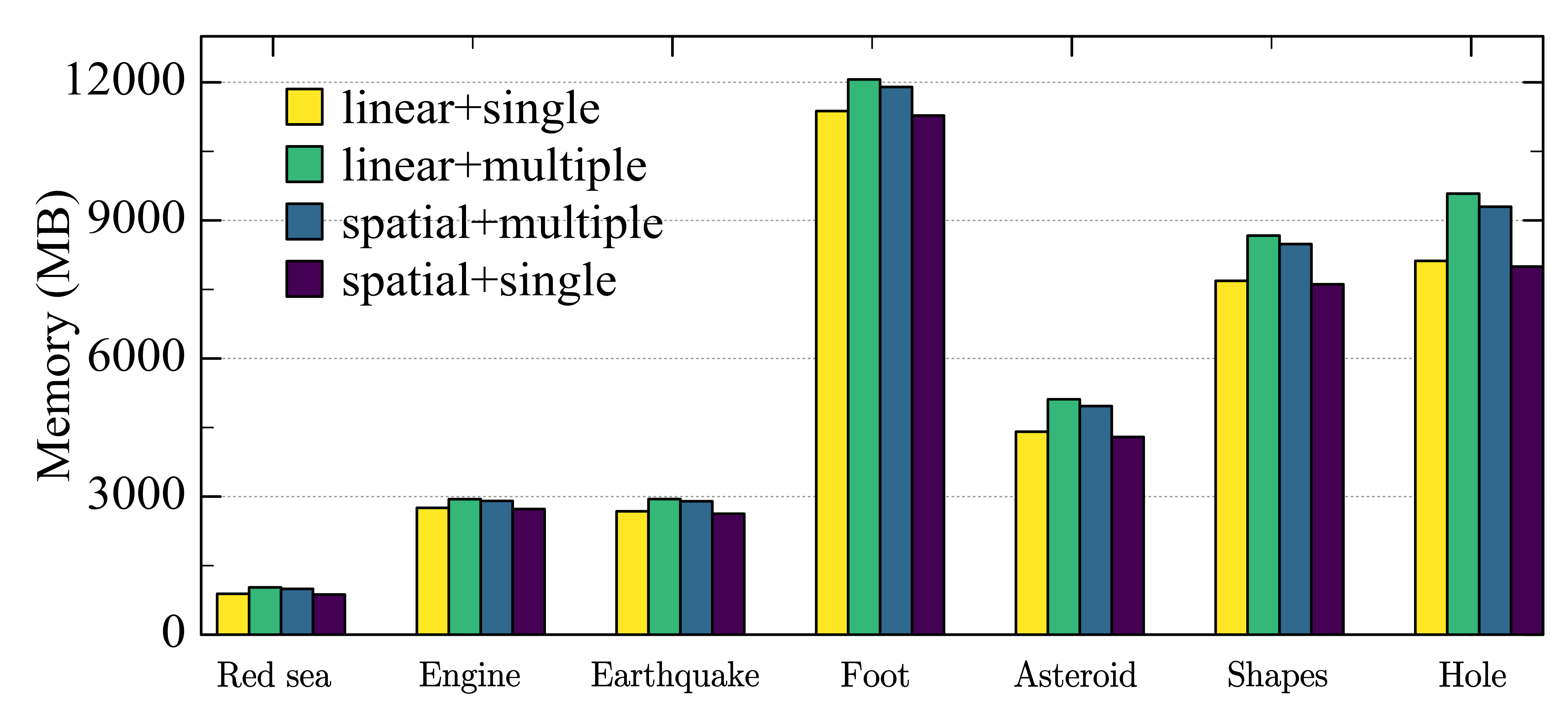} \\
    (a) & (b) \\ 
    \end{tabular}
    \caption{The execution time (in seconds) and memory usage (in megabytes) used by \texttt{Morse\-Smale\-Complex} plugin when running with different computing modes.}
    \label{fig:res_mode_ms}
\end{figure}

The opposite is true for \texttt{Scalar\-field\-Critical\-Points} (\cref{fig:res_mode}(b)) and \texttt{Discrete\-Gradient} plugins (\cref{fig:res_mode}(c)). Although differences across modalities are less pronounced, computing all listed topological relations using the linear buffer results in algorithms running 14\% faster. This is reasonable since the plugin uses multiple relations for each vertex at a time. 
Not surprisingly, precomputing only one single topological relation also works better for the \texttt{Morse\-Smale\-Complex} plugin (\cref{fig:res_mode_ms}(a)). Unlike the other two plugins, this plugin utilizes topological relations to reconstruct Morse-Smale cells while navigating the entire mesh. Due to the navigation process only focusing on one topological relation at a time, the single mode is beneficial. Moreover, the spatial buffer provides slightly better performance than the linear one. This is because the computation of the MS complex follows the path according to a discrete gradient field and does not always visit blocks in a linear order. The spatial buffer precomputing a single relation is 85\% faster on average.

Memory usage does not change significantly across modalities. As we can expect, the work modes precomputing multiple relations use more space compared to those precomputing only one relation, up to 18\% as observed in our experiments. Overall memory usage grows linearly with the number of tetrahedra in the input mesh. We include detailed graphs in the additional materials only. The only exception is the \texttt{Morse\-Smale\-Complex} plugin (\cref{fig:res_mode_ms}). For this plugin, we can also notice that the time and memory do not increase consistently as the size of datasets increases; this is because extracting the MS cells depends on the size of the MS complex instead of the input mesh (i.e., output-sensitive). Still, computing multiple relations for each block requires about 12\% more memory space than computing a single relation.

\textbf{Lessons learned.} Overall, the results match our hypothesis. Specifically, the computing mode of linear buffer with the single relation works best for \texttt{Test\-Topo\-Relations} plugin, the mode of linear buffer with all required relations performs best for \texttt{Scalar\-Field\-Critical\-Points} and \texttt{Discrete\-Gradient} plugins, and the mode of spatial buffer with single relation is optimal for \texttt{Morse\-Smale\-Complex} plugin. Furthermore, using a spatial buffer with computing all required relations is not optimal for any of the testing plugins.

\subsubsection{Comparison of different numbers of producers}
\label{sec:results_thread}

The number of producer threads is another parameter that affects performance. Our hypothesis is that more producer threads should accelerate the execution of the plugin by reducing the waiting time of the consumer thread. For the following experiments, we adopted the best computing mode for each plugin, as observed in \cref{sec:results_mode}, while changing the number of producer threads from 1 to 10. Given that the same computing mode and buffer capacity are used, differences in memory costs are very limited (with a maximum increase of 16\% for \texttt{Discrete\-Gradient} plugin and less than 1\% for all others). Therefore, we only discuss time performance in the following, while providing detailed graphs in supplemental materials.

\cref{fig:res_thread} shows the time performance of all four tested plugins. For all of them, we can notice that using more producers reduces the waiting time of the consumer thread. For all plugins, except the \texttt{Morse\-Smale\-Complex} plugin and \texttt{Scalar\-Field\-Critical\-Points} plugin on the Stent dataset, the waiting time disappears after using 6 producers (as indicated by the cyan colored bar in the figure). Overall, the speedup plateaus at around 6 producers, after which producers will end up competing with the consumer for computing resource (i.e., which thread to be scheduled), and the context switching adds overhead to the performance. Using 6 threads results in a 4.2 times speedup if we consider the extraction of topological relations only (\texttt{Test\-Topo\-Relations}), and about 1.7 times speedup for all remaining plugins.

\begin{figure}[htb!]
    \begin{tabular}{cc}
        \includegraphics[width=0.45\linewidth]{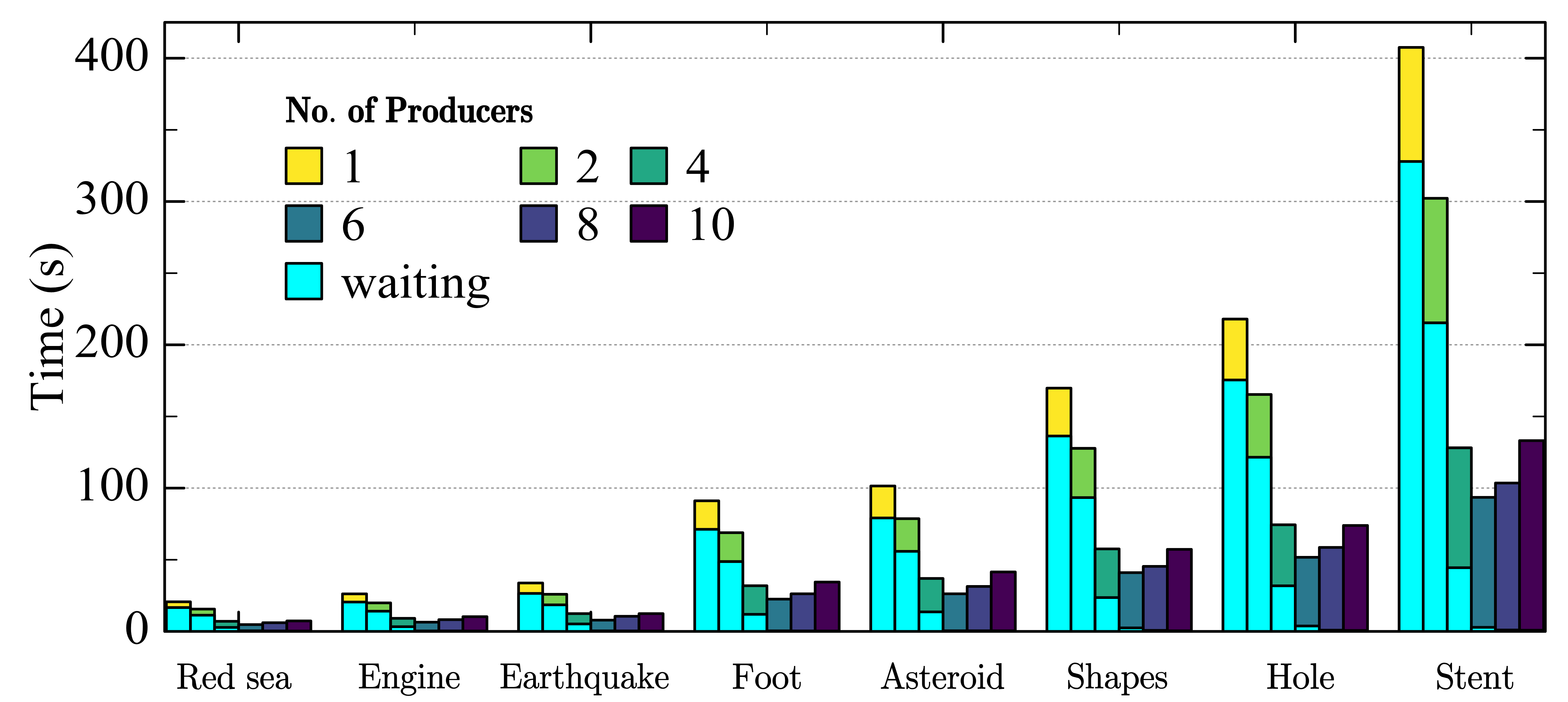} & 
        \includegraphics[width=0.45\linewidth]{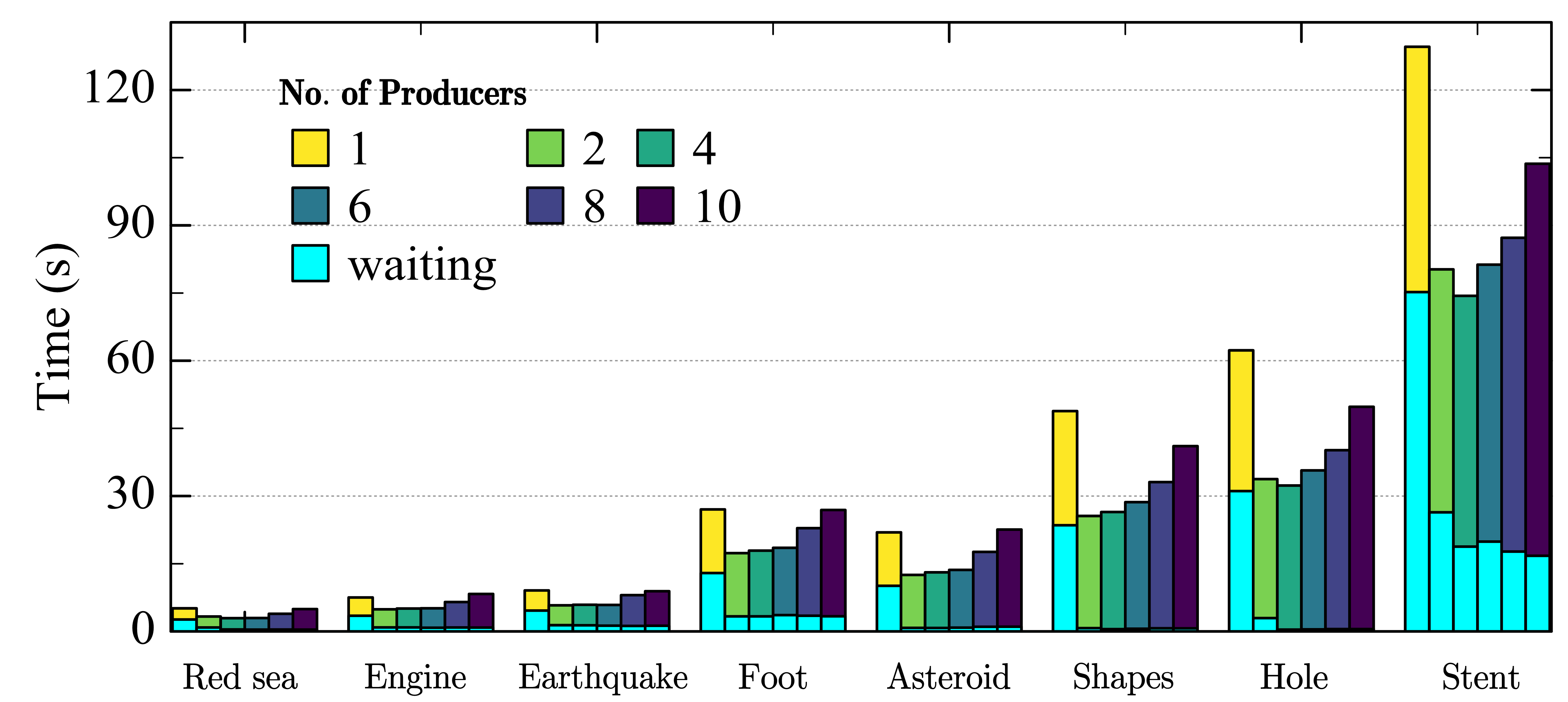} \\
        (a) \small \texttt{Test\-Topo\-Relations} & 
        (b) \small \texttt{Scalar\-Field\-Critical\-Points} \\ \\
        \includegraphics[width=0.45\linewidth]{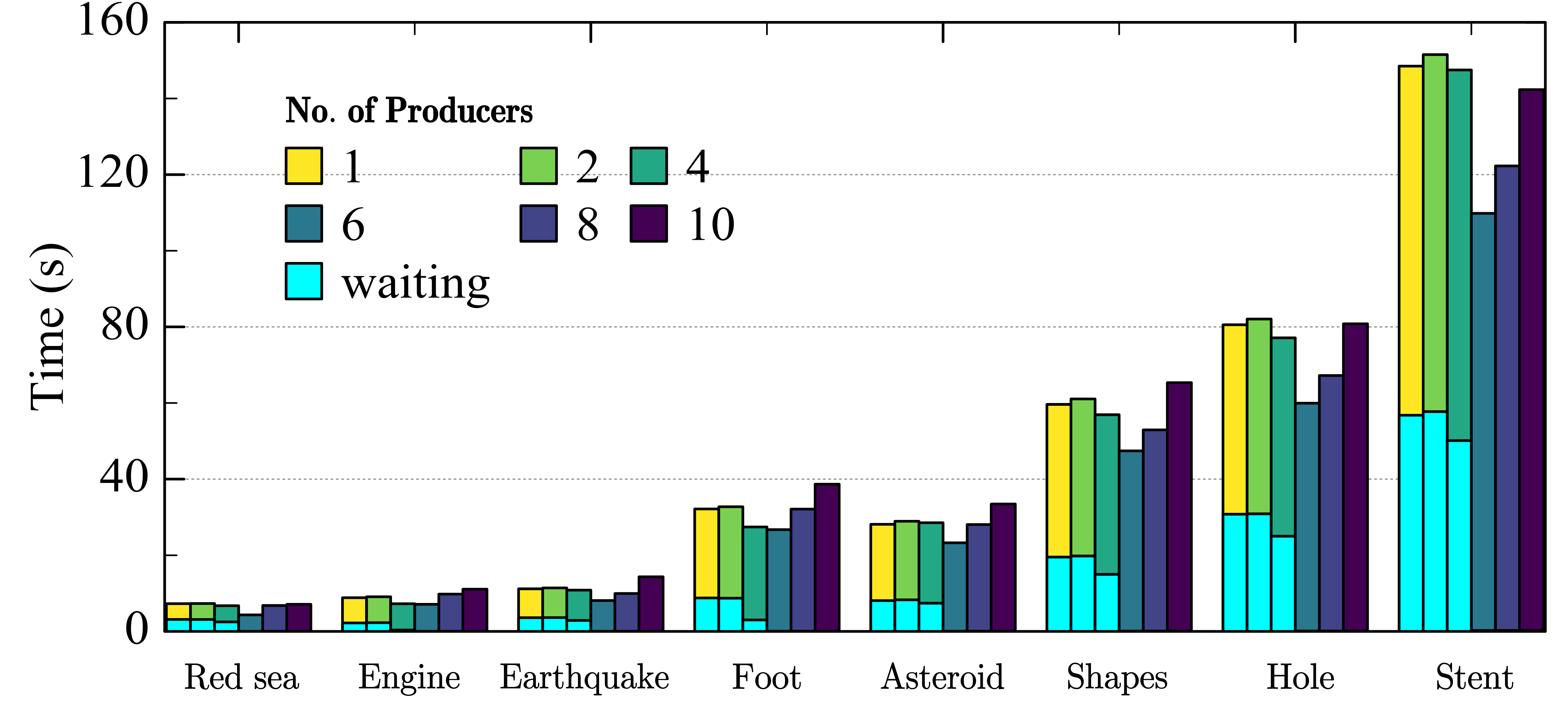} & 
        \includegraphics[width=0.45\linewidth]{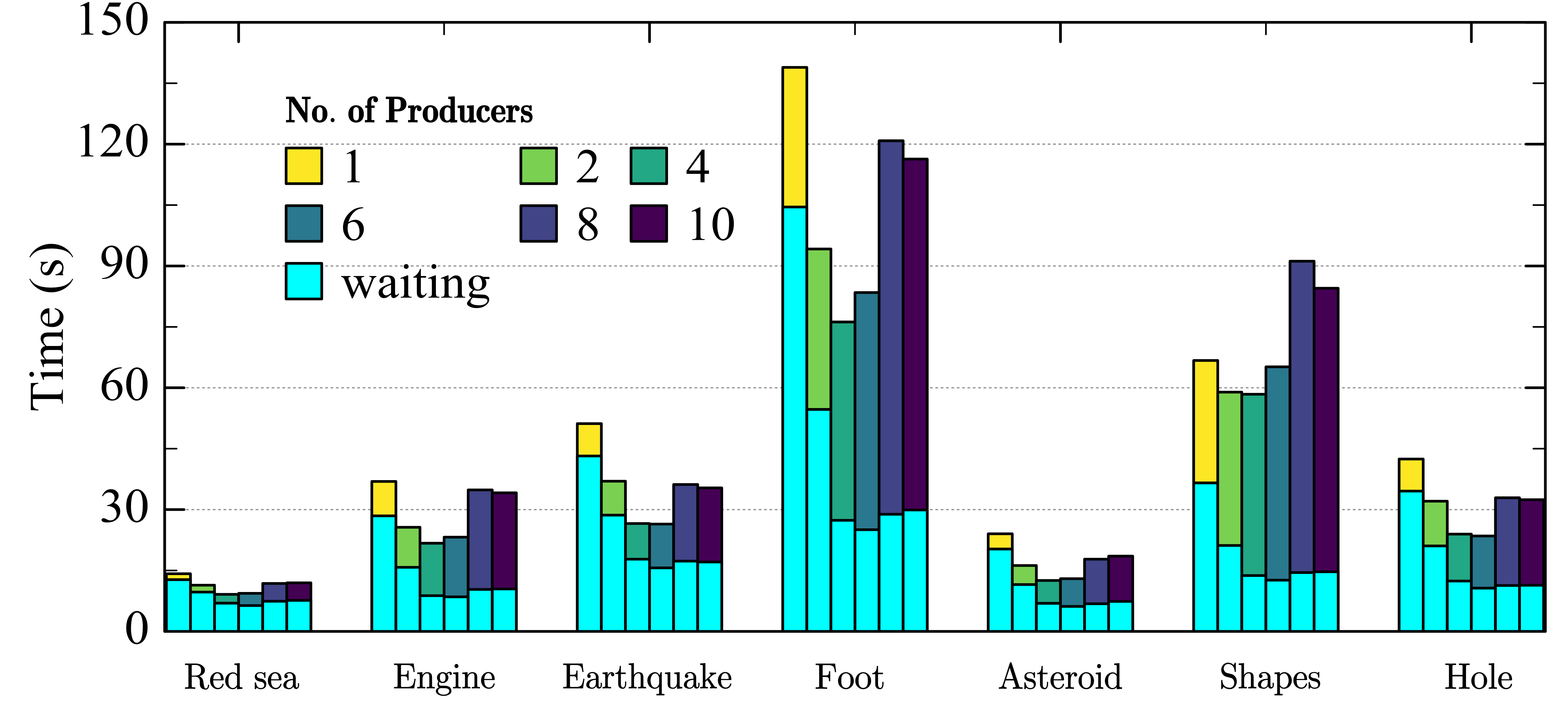} \\
        (c) \small \texttt{Discrete\-Gradient} & 
        (d) \small \texttt{Morse\-Smale\-Complex} \\
    \end{tabular}
    \caption{The total time (in seconds) spent by four testing plugins when running with different numbers of producer threads, and the cyan colored bar denotes the waiting time of the consumer thread.}
    \label{fig:res_thread}
\end{figure}


\textbf{Lessons learned.} Overall, the results show that multiple producer threads can improve the time performance of the plugin, but such speedup is limited. For most testing plugins, using 4 producer threads shows significant improvements over the single producer. However, configuring the number of producer threads to exceed the number of physical cores will result in performance degradation. This is expected due to issues such as operating system's scheduling policy, context switching, and critical resource competition.

\subsection{Comparing with state-of-the-art}

In this section, we compare our proposed data structure with state-of-the-art static and dynamic data structures. For static data structure, we use {\em ExplicitTriangulation} implemented in TTK. This data structure precomputes and stores all required topological relations during a preprocessing stage so that they are readily available during the algorithm execution. For dynamic data structure, we use {\em CompactTriangulation}, an implementation of the dynamic data structure proposed by Liu et al. \cite{Liu2021topocluster}. Similar to ACTOPO, CompactTriangulation organizes the mesh using blocks, computes and discards topological relations locally to each block.

The comparison is intended to evaluate the parallelism at the data structure level instead of the algorithm level. For this reason, the testing algorithms are restricted to the sequential execution.
In practice, multiple threads are used for initializing the data structures or, in the case of ACTOPO, threads are used for supporting the extraction of topological relations (6 producers). Instructions implemented by each plugin are executed by a single thread (1 consumer for ACTOPO) in a sequential manner.

Since the initialization of each data structure is different, we report detailed results about the preprocessing time (steps required to initialize the data structure) and the algorithm execution time. For the following experiments, we have set the number of threads to 6 for all data structures and limited the buffer/cache capacity to 20\% of the total number of blocks for both our proposed data structure and TTK CompactTriangulation. 

\begin{figure*}[!htb]
    \centering
    \newcolumntype{A}{>{\centering\arraybackslash} m{0.3\linewidth}}
    \begin{tabular}{A A A}
        \toprule
        \texttt{Test\-Topo\-Relations} & \texttt{Scalar\-Field\-Critical\-Points} & \texttt{Discrete\-Gradient}\\
        \midrule \\ 
        \includegraphics[width=\linewidth]{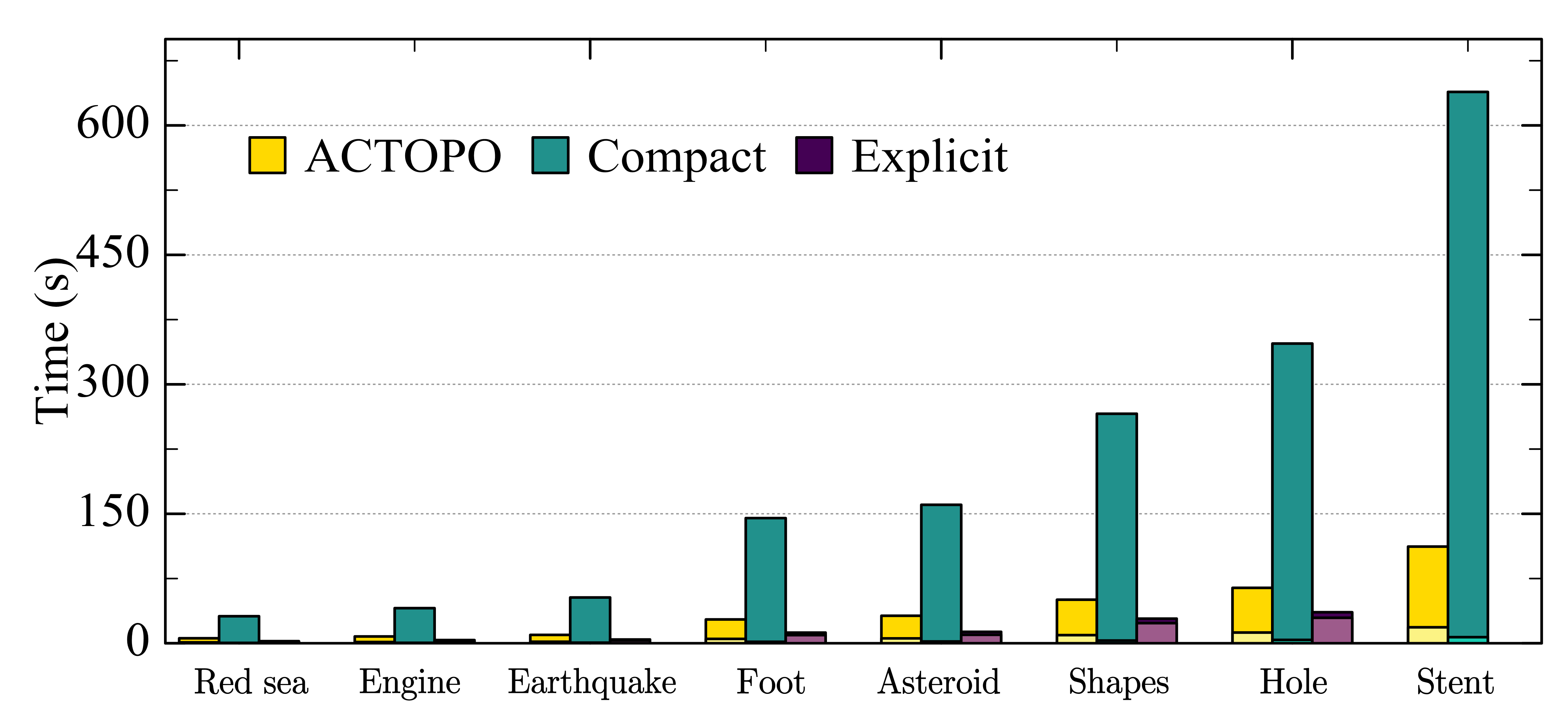} & 
        \includegraphics[width=\linewidth]{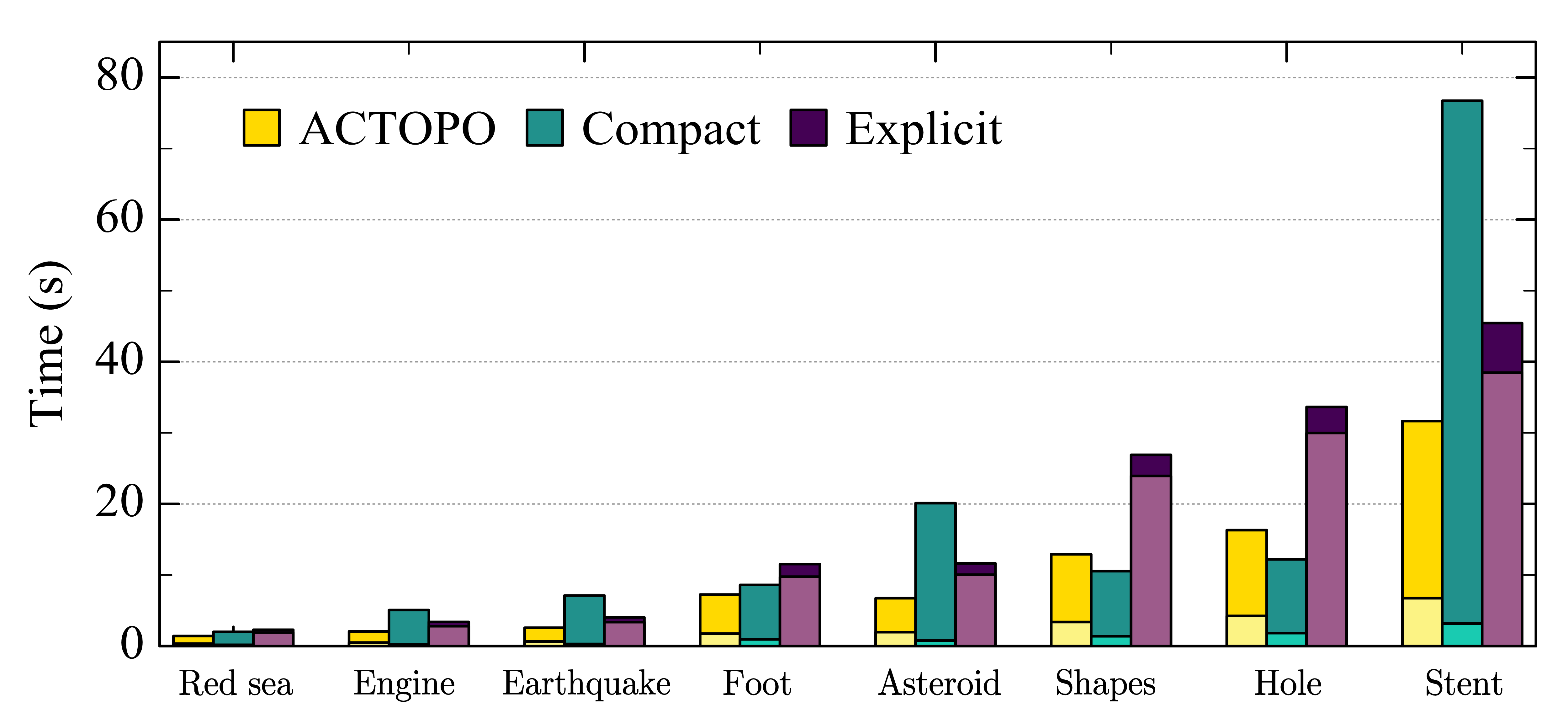} & 
        \includegraphics[width=\linewidth]{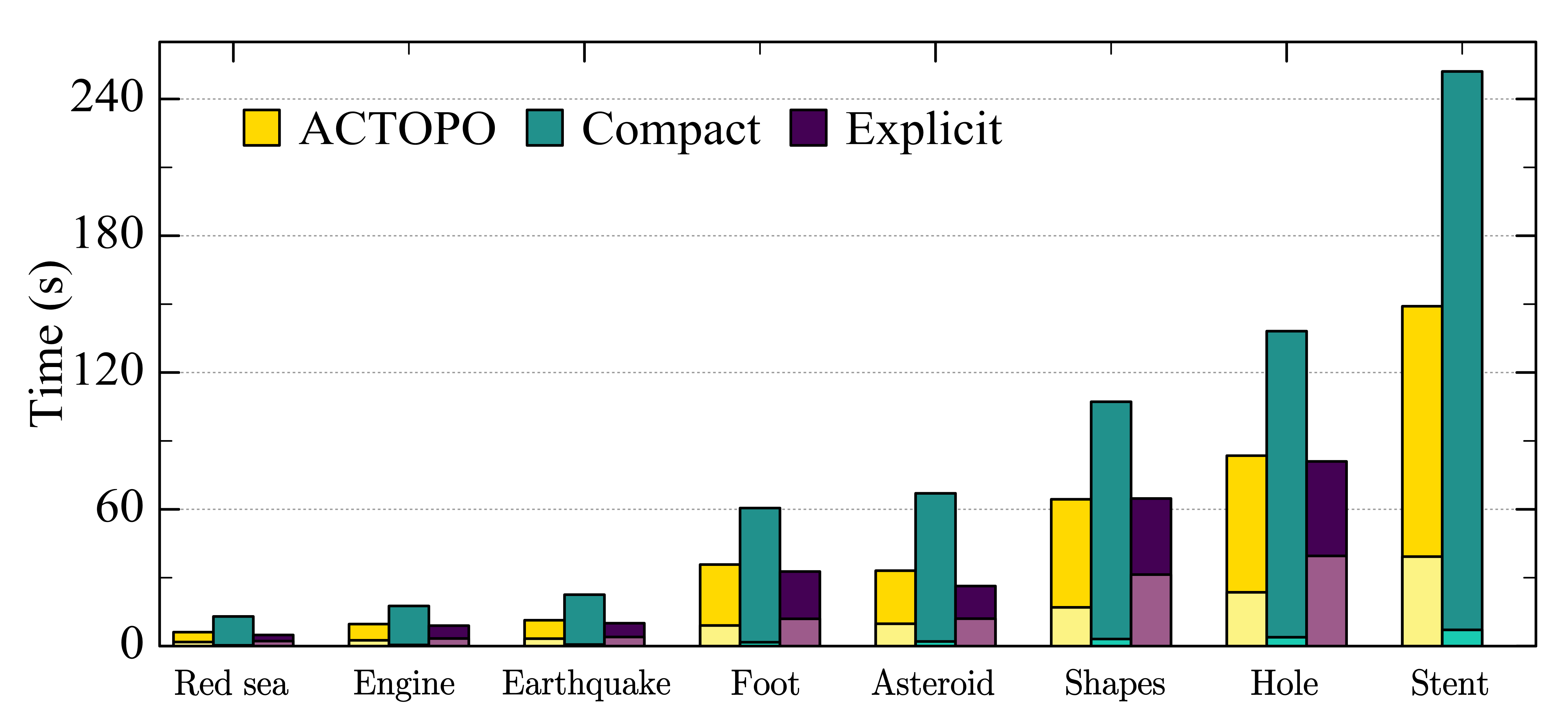} \\
        (a) & (c) & (e) \\
        
        \includegraphics[width=\linewidth]{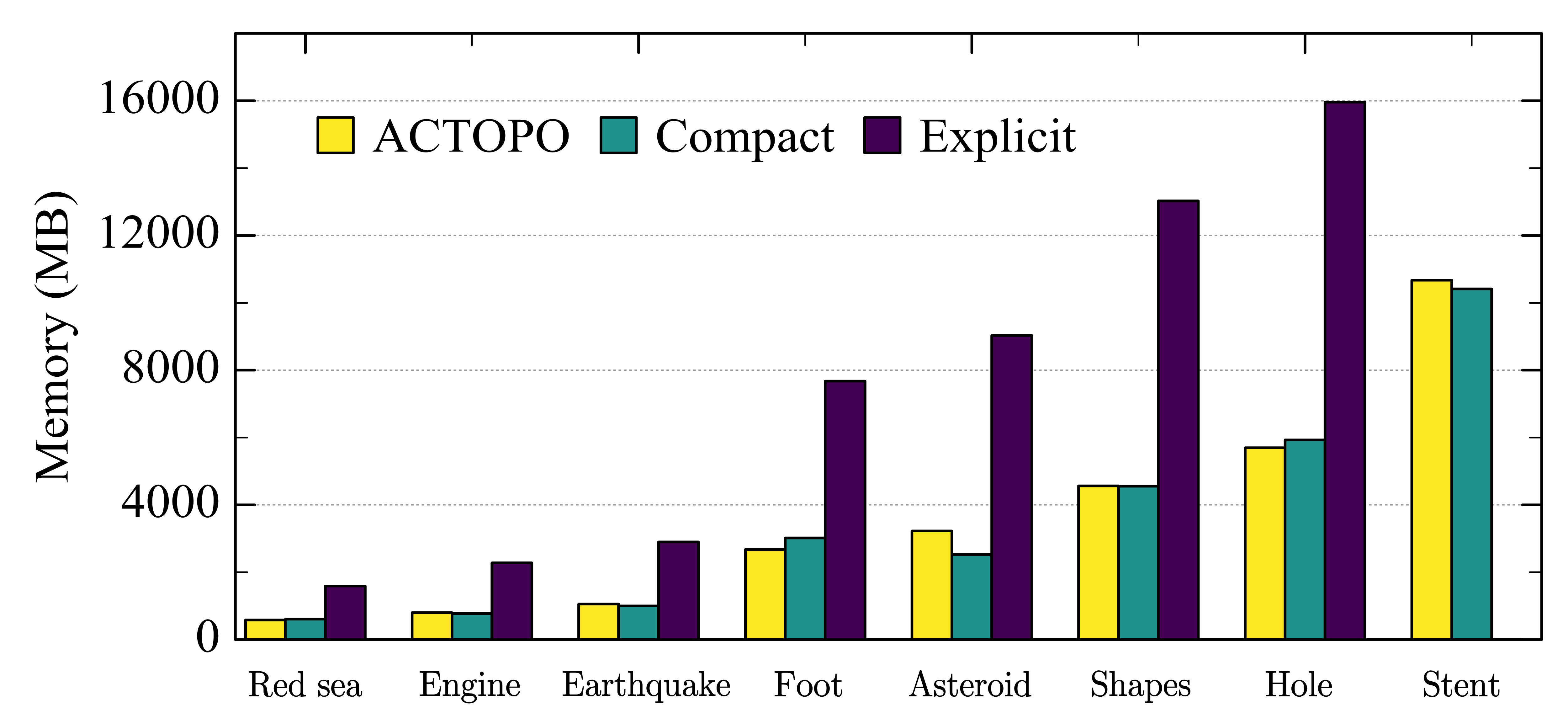} &
        \includegraphics[width=\linewidth]{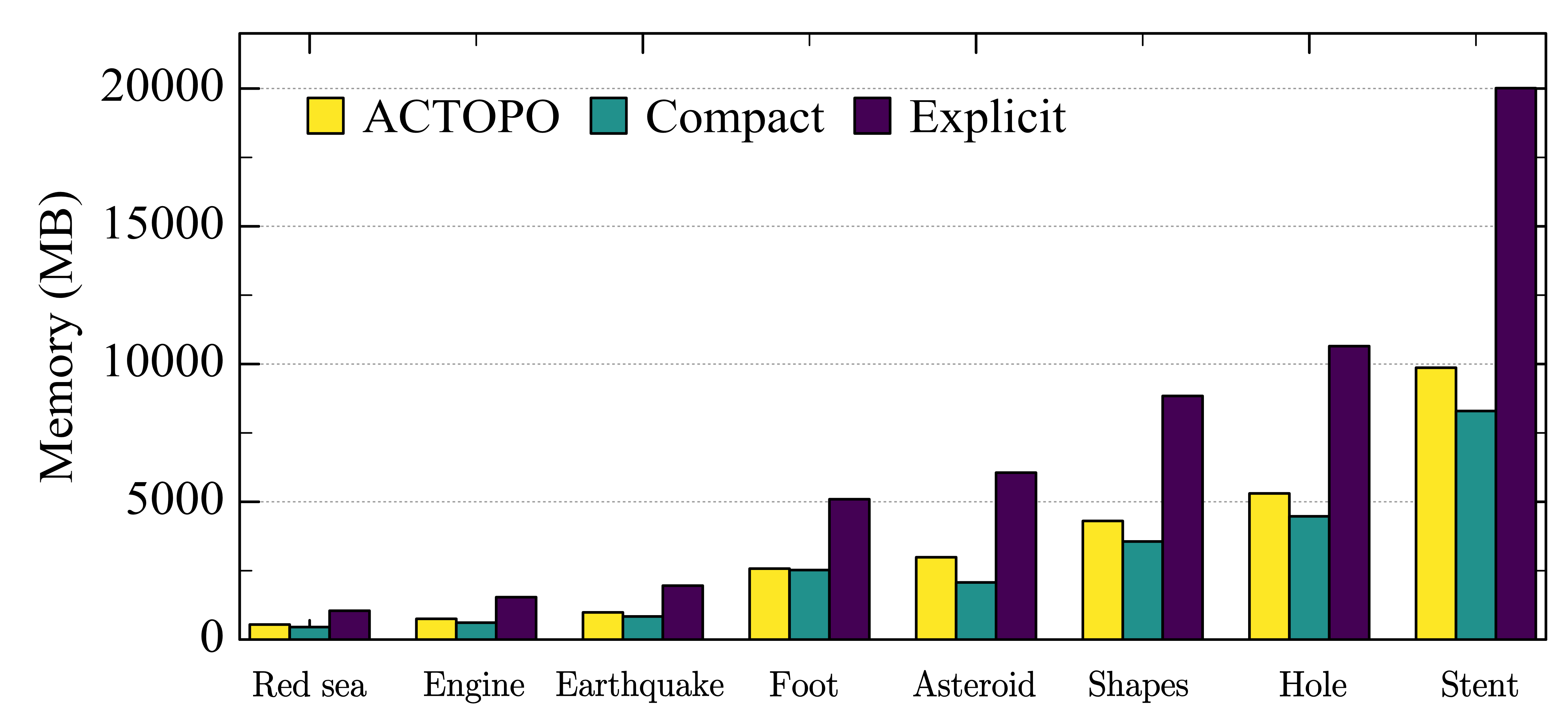} &
        \includegraphics[width=\linewidth]{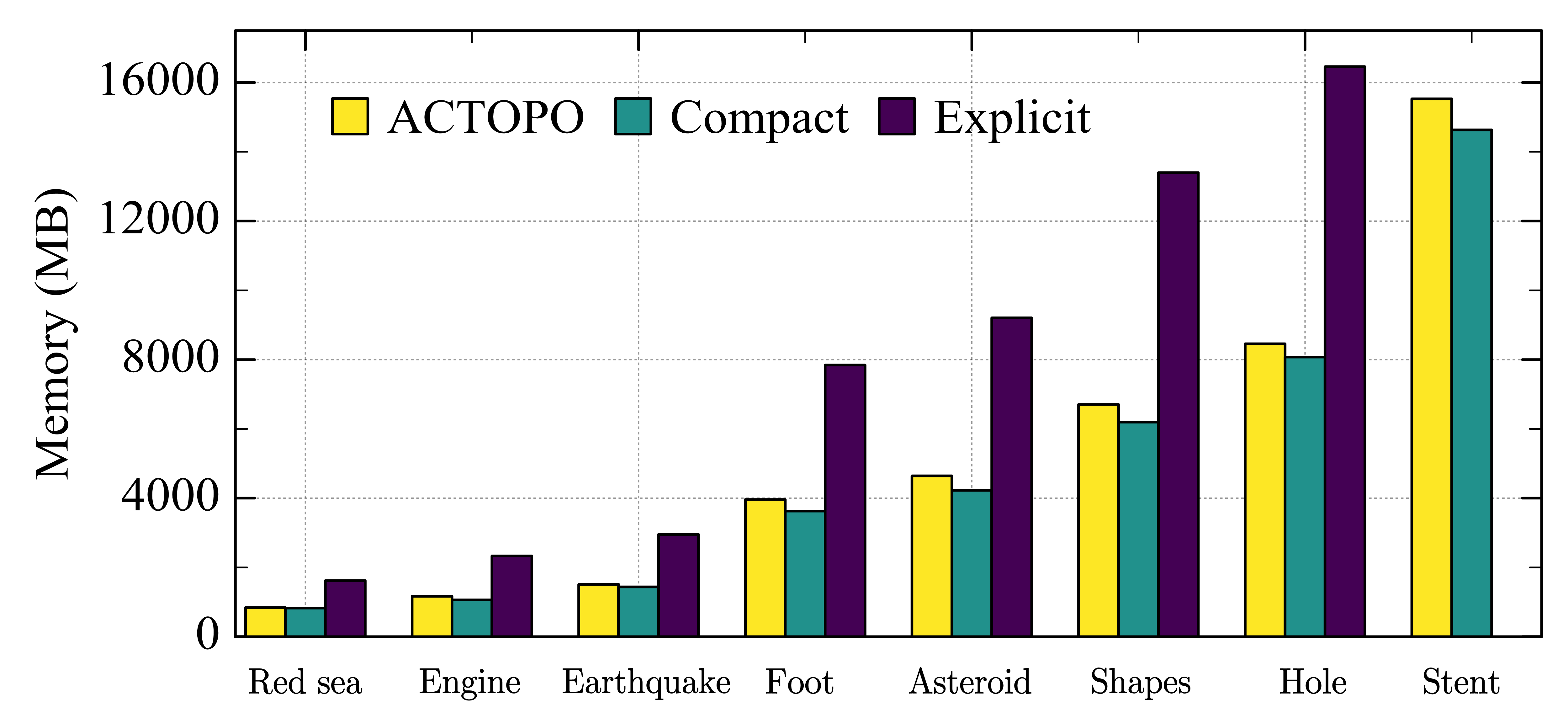} \\
        (b) & (d) & (f) \\
        \bottomrule
    \end{tabular}
    \caption{The total time (in seconds) and memory usage (in megabytes) used by three experimental plugins with different data structures: ACTOPO, TTK Compact Triangulation, and TTK Explicit Triangulation when the algorithm runs in sequential. In (a), (c), and (e), the top bar indicates the time required by running the algorithm, while the bottom bar indicates the preprocessing time.}
    \label{fig:res_cmp}
\end{figure*}

\cref{fig:res_cmp} shows the total time and memory consumptions obtained with three linear plugins. Focusing on the preprocessing step, we notice that our proposed data structure is the second best on average in terms of time performance. {\em ExplicitTriangulation} is the slowest being in general, 2 times as slow as ACTOPO, and {\em CompactTriangulation} is 4 times as fast as ACTOPO. For the \texttt{Test\-Topo\-Relations} plugin (\cref{fig:res_cmp}(a)), CompactTriangulation shows the worst performance instead, underlying the limitation of computing topological relations on-the-fly while the algorithm is blocked to wait. Our approach provides a dramatic improvement and is 5.3 times as fast as CompactTriangulation.


In the remaining two plugins (\cref{fig:res_cmp}(c) and (e)), topological relations are used for computing additional information. Producer threads used in our approach have enough time to precompute topological relations and save time for the consumer. This allows ACTOPO to close the gap with ExplicitTriangulation and to show roughly the similar performance. The difference with CompactTriangulation is still significant, since ACTOPO is 3.4 times as fast as CompactTriangulation for \texttt{Scalar\-Field\-Critical\-Points} plugin and 1.83 times for \texttt{Discrete\-Gradient} plugin.

In terms of memory usage, even if {\em CompactTriangulation} is the most compact data structure in general, ACTOPO shows a very similar memory footprint using only 7\% more memory on average. As opposed, ExplitTriangulation uses about 3 times the memory of our proposed data structure. The large memory requirement causes ExplicitTriangulation to run out of memory when processing the Stent dataset with \texttt{Test\-Topo\-Relations} plugin.


\begin{figure}[!htb]
    \centering
    \begin{tabular}{cc}
    \includegraphics[width=0.45\linewidth]{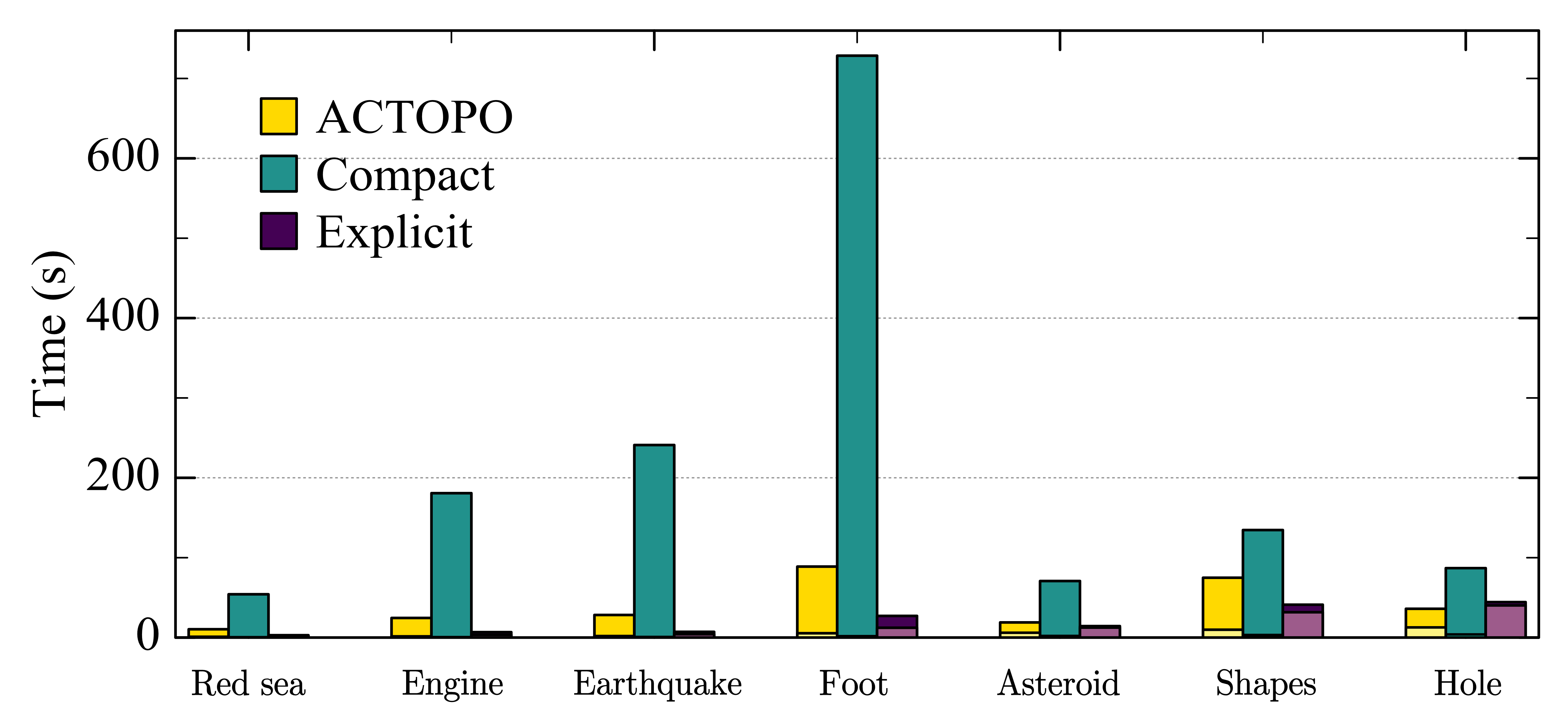} & 
    \includegraphics[width=0.45\linewidth]{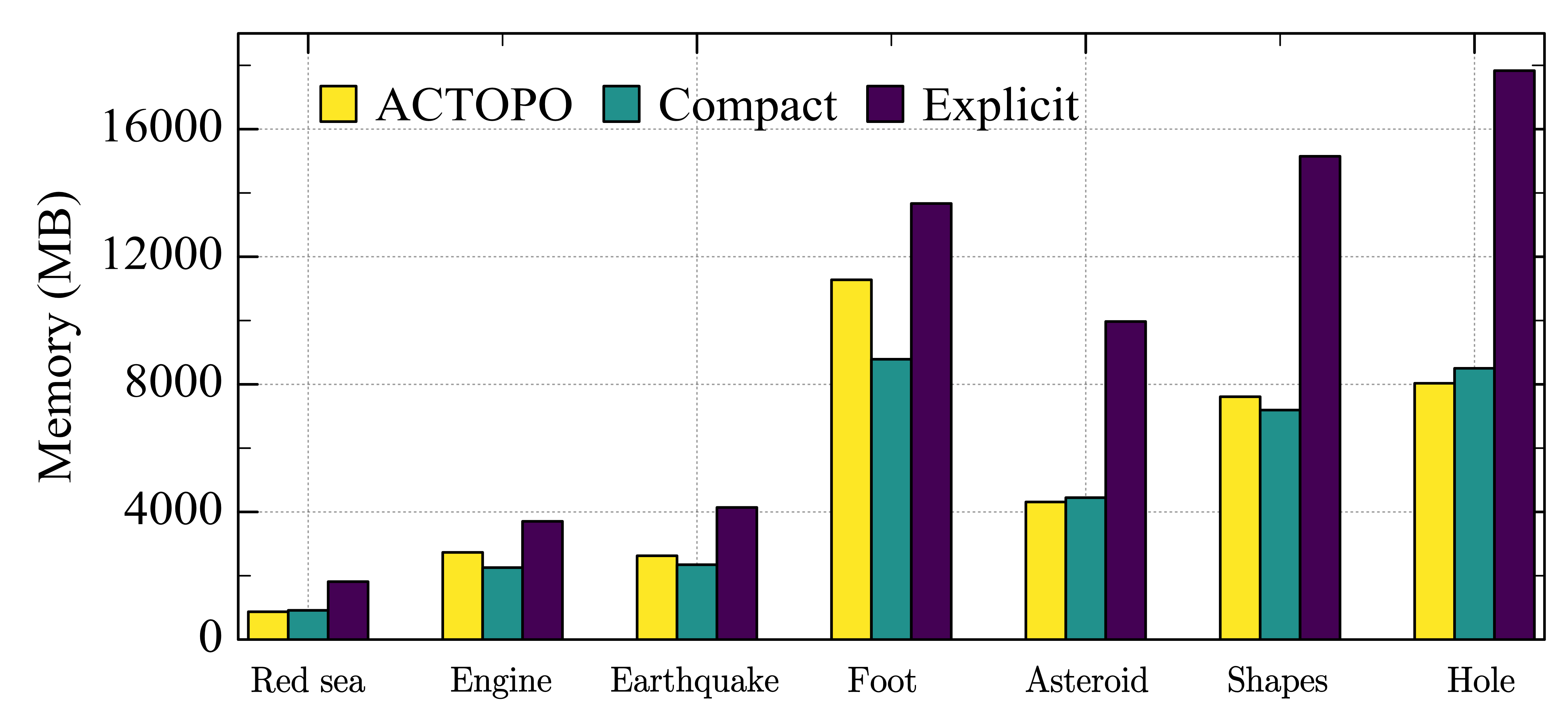} \\
    (a) & (b) \\
    \end{tabular}
    \caption{The total time (in seconds) and memory usage (in megabytes) used by \texttt{Morse\-Smale\-Complex} plugin when running with different data structures: ACTOPO, TTK Compact Triangulation, and TTK Explicit Triangulation.}
    \label{fig:res_cmp_ms}
\end{figure}

Particular attention is dedicated to the results collected on the \texttt{Morse\-Smale\-Complex} plugin, shown in \cref{fig:res_cmp_ms}. Since the plugin does not simply iterate through the simplices of the mesh, block-based data structures like CompactTriangulation are forced to recompute topological relations inside a block multiple times with an evident loss in performance, especially when handling the input dataset with a complicated scalar field, such as Engine, Earthquake, and Foot datasets. ACTOPO provides a considerable improvement being 3 times as fast as CompactTriangulation. However, the unpredictable pattern used by the plugin to access blocks makes it harder for the producers to predict which block the consumer will move to next. As a result, the gap between ACTOPO and ExplicitTriangulation is wider with ACTOPO being 2.6 times as slow as ExplicitTriangulation on average. Nevertheless, this is paid off by much better scalability in terms of memory usage, with ACTOPO being 1.8 times as compact as ExplicitTriangulation. As mentioned earlier, the plugin is output-sensitive, which causes both localized data structures to use the most memory when processing the Foot dataset. 

\section{Supporting parallel algorithms}
\label{sec:para_algorithm}

Parallel computation makes use of multiple processors to execute computational tasks simultaneously, thus reducing the total time required to complete the task. Various topology-based visualization algorithms have implemented different parallel computation techniques, and one common approach is the data-parallel technique. In this method, data is divided into smaller pieces and processed independently by multiple threads. In this section, we describe how ACTOPO supports a parallel topological algorithm.  

For static data structures, data-parallel techniques are used to parallelize the computation of topological relations, or the selected algorithm, independently. For dynamic data structures, data-parallel techniques are used to allow multiple threads to process multiple blocks at the same time \cite{Liu2021topocluster}. 

Our proposed ACTOPO data structure can also be adapted to execute parallel algorithms. As mentioned previously, the consumer thread is the one responsible for running the processing algorithm. When an algorithm allows for parallel execution, multiple consumer threads can be used to run instructions in parallel. To this end, we have generalized our model by duplicating producer-consumer associations.

Consumer threads are spawned at the initialization phase of the ACTOPO. Once created, each consumer thread has its own set of producers and a dedicated buffer system that is not shared with any other consumer/producer threads. Each type of thread behaves in the same way as described in \cref{sec:seq_algorithm}, i.e., the consumer thread is responsible for the algorithm execution, the worker producer threads precompute topological relations and saves them to the buffer storage, and the leader producer thread monitors both threads and manages the buffer system. The only difference is the exclusive association between consumers and producers.

In the following, we present an evaluation of this producer-consumer paradigm when the topological algorithm runs in parallel. We keep only three of the four TTK plugins originally selected since \texttt{Morse\-Smale\-Complex} uses parallelism only for computing the discrete gradient (equivalent to the \texttt{Discrete\-Gradient} plugin), while the extraction of the MS cells is done in a sequential manner.

\subsection{Evaluating robustness on parallel algorithms}

When working with a parallel algorithm, the total number of consumer and producer threads involved is an important parameter to evaluate. The total number of threads used by ACTOPO is $t_c + t_c \cdot t_{pc}$, where $t_c$ indicates the number of consumer threads, $t_{pc}$ indicates the number of producer threads per consumer thread, including one leader producer and ($t_{pc}-1$) worker producers. Thus, each plugin can obtain a relative speedup by either increasing $t_c$ to distribute the computation across more consumer threads, or by increasing $t_{pc}$ to let more worker threads precompute topological relations. Since our testing machine can support at most 12 threads, we can use at least 2 consumers with 5 producers each (i.e., $t_c=2$ and $t_{pc}=5$), and at most 6 consumers with 1 (leader) producer assigned to each one of them (i.e., $t_c=6$ and $t_{pc}=1$).

\begin{figure*}[!htb]
    \centering
    \newcolumntype{A}{>{\centering\arraybackslash} m{0.3\linewidth}}
    \begin{tabular}{A A A}
        \toprule
        \texttt{Test\-Topo\-Relations} & \texttt{Scalar\-Field\-Critical\-Points} & \texttt{Discrete\-Gradient}\\
        \midrule \\ 
        \includegraphics[width=\linewidth]{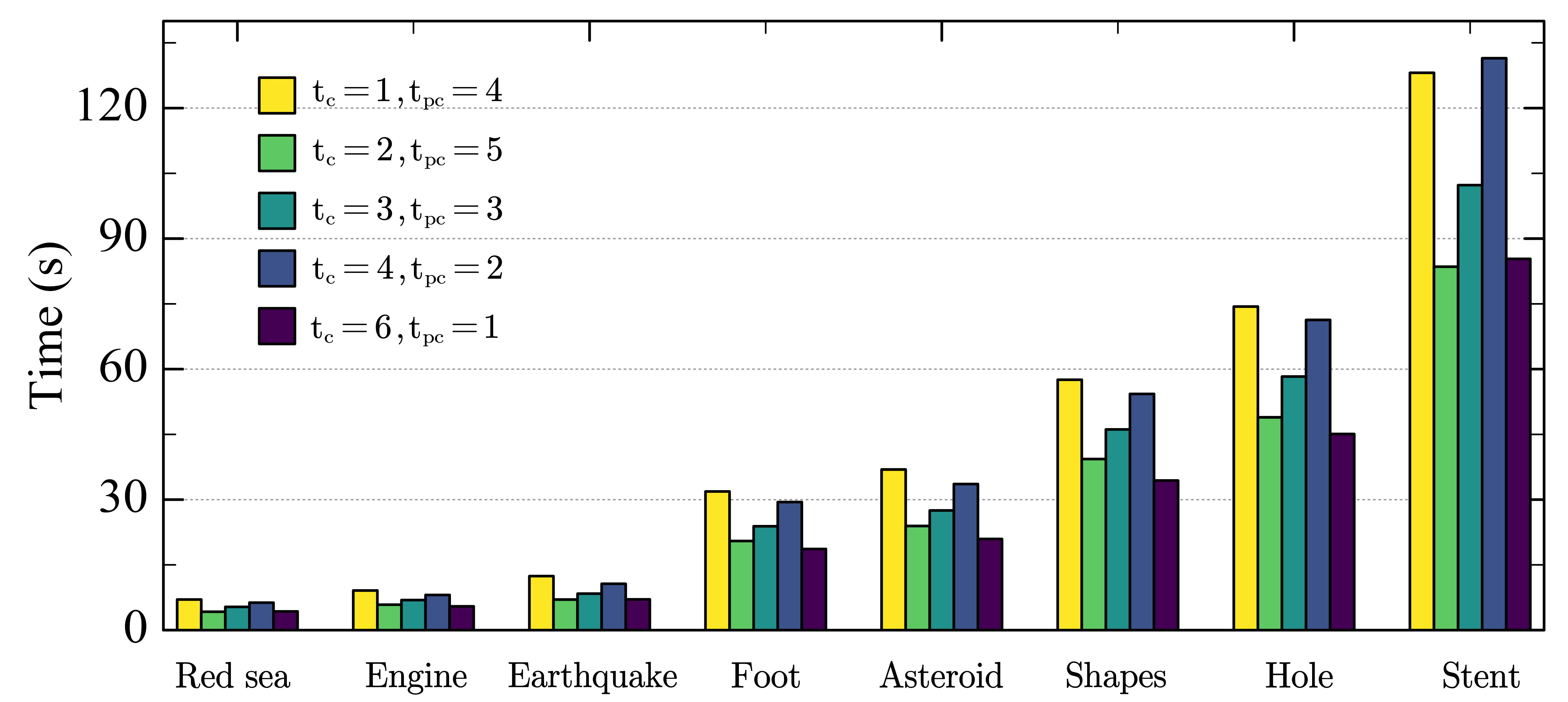} & 
        \includegraphics[width=\linewidth]{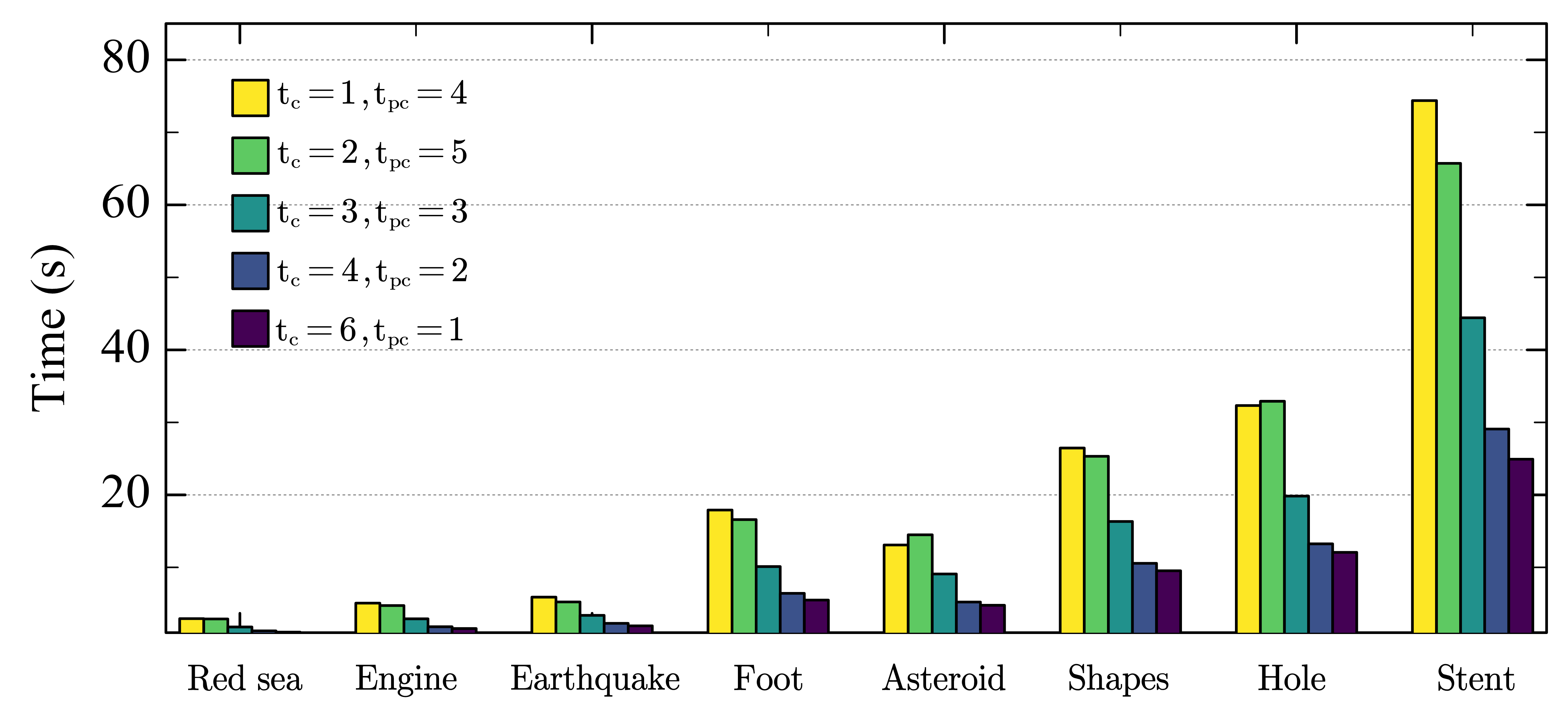} & 
        \includegraphics[width=\linewidth]{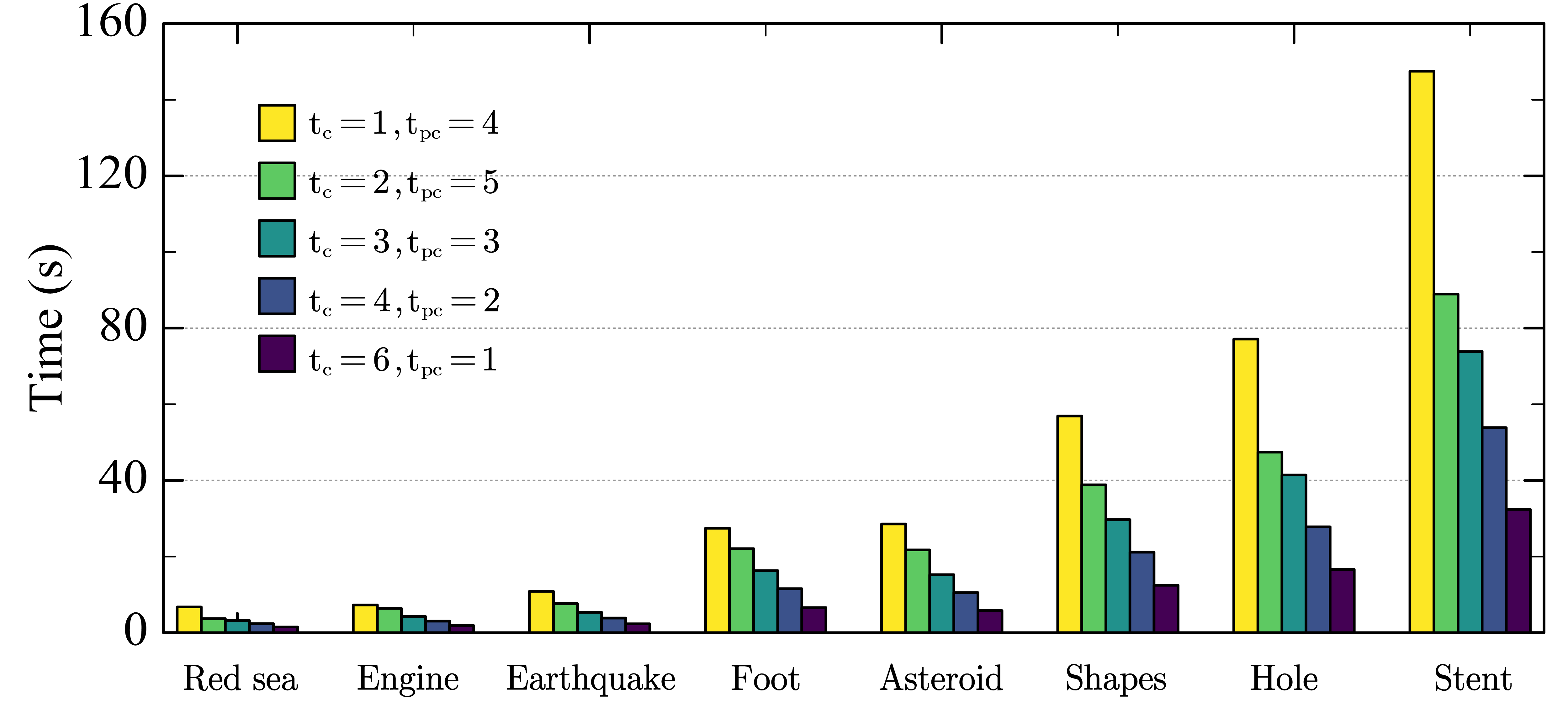} \\
        (a) & (b) & (c) \\
        \bottomrule
    \end{tabular}
    \caption{The total time (in seconds) used by three experimental plugins with different numbers of consumer and producer threads when the algorithm runs in parallel with OpenMP.}
    \label{fig:res_consumer_producer_num}
\end{figure*}

\cref{fig:res_consumer_producer_num} shows the total time and peak memory tracked for three testing plugins. We also reuse the results obtained with a single consumer thread (sequential computation) in \cref{sec:results_thread} as a reference. 


In general, using the maximum number of consumer threads achieves the best time performance. Since the memory usages are at a similar level (within 3\% difference) except the \texttt{Discrete\-Gradient} plugin (within 10\% difference), we only discuss the time performance here and shared the detailed figures in supplemental materials. 

The \texttt{Test\-Topo\-Relations} plugin only extracts the topological relations from the mesh without computing any additional information, the precomputation from five producer threads helps speed up the execution, but it is still on average 5\% slower than fully distributing the task to consumers. With one more consumer thread, it is 1.7 times as fast as the single consumer thread.

The \texttt{Scalar\-Field\-Critical\-Points} and \texttt{Discrete\-Gradient} plugins show the advantage of maximizing the number of consumer threads over other cases, and it is on average 2.4 times as fast as maximizing the number of producer threads for both plugins.

\textbf{Lessons learned}. If the parallel processing algorithm is intensive with topological relations, i.e., the use of topological relations is faster than the production, both maximizing the consumer and producer threads will speed up the execution. Otherwise, it is preferred to increase the number of consumer threads to distribute the workload instead of increasing the number of producer threads for precomputation. Even though more work producers can reduce the waiting time of the consumer, more consumer threads can make the algorithm proceed in parallel and thus reduce the total time spent by the algorithm.

\subsection{Comparing with state-of-the-art}

In this section, we repeat our comparison with CompacTriangulation and ExplicitTriangulation data structures, allowing testing plugins to run in parallel. In the following experiments, all data structures use 12 threads for their initialization (preprocessing). CompacTriangulation and ExplicitTriangulation use 12 threads to run the requested algorithm. Our data structure uses 6 consumer threads supported by 1 leader producer each (i.e., $t_c=6$ and $t_{pc}=1$).

\begin{figure*}[!htb]
    \centering
    \newcolumntype{A}{>{\centering\arraybackslash} m{0.3\linewidth}}
    \begin{tabular}{A A A}
        \toprule
        \texttt{Test\-Topo\-Relations} & \texttt{Scalar\-Field\-Critical\-Points} & \texttt{Discrete\-Gradient}\\
        \midrule \\ 
        \includegraphics[width=\linewidth]{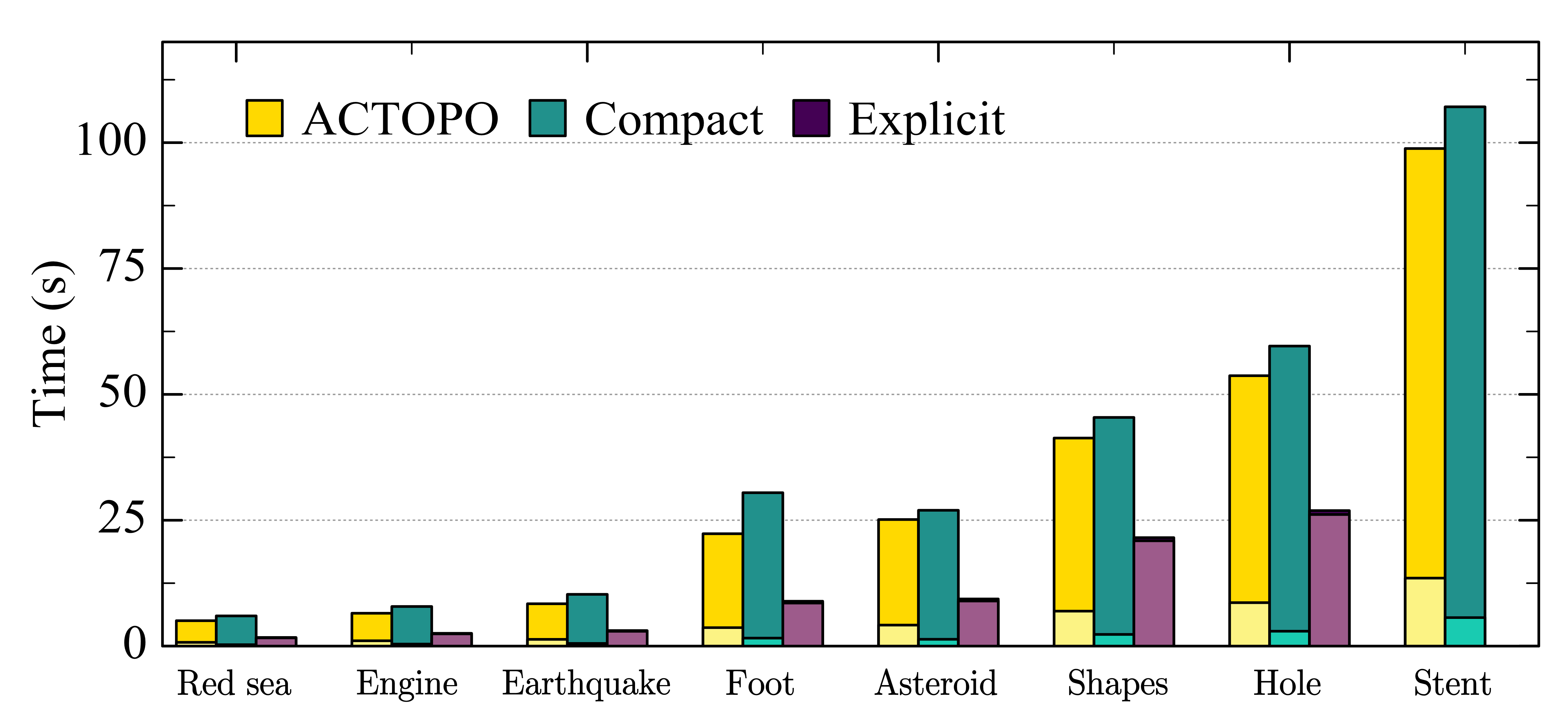} & 
        \includegraphics[width=\linewidth]{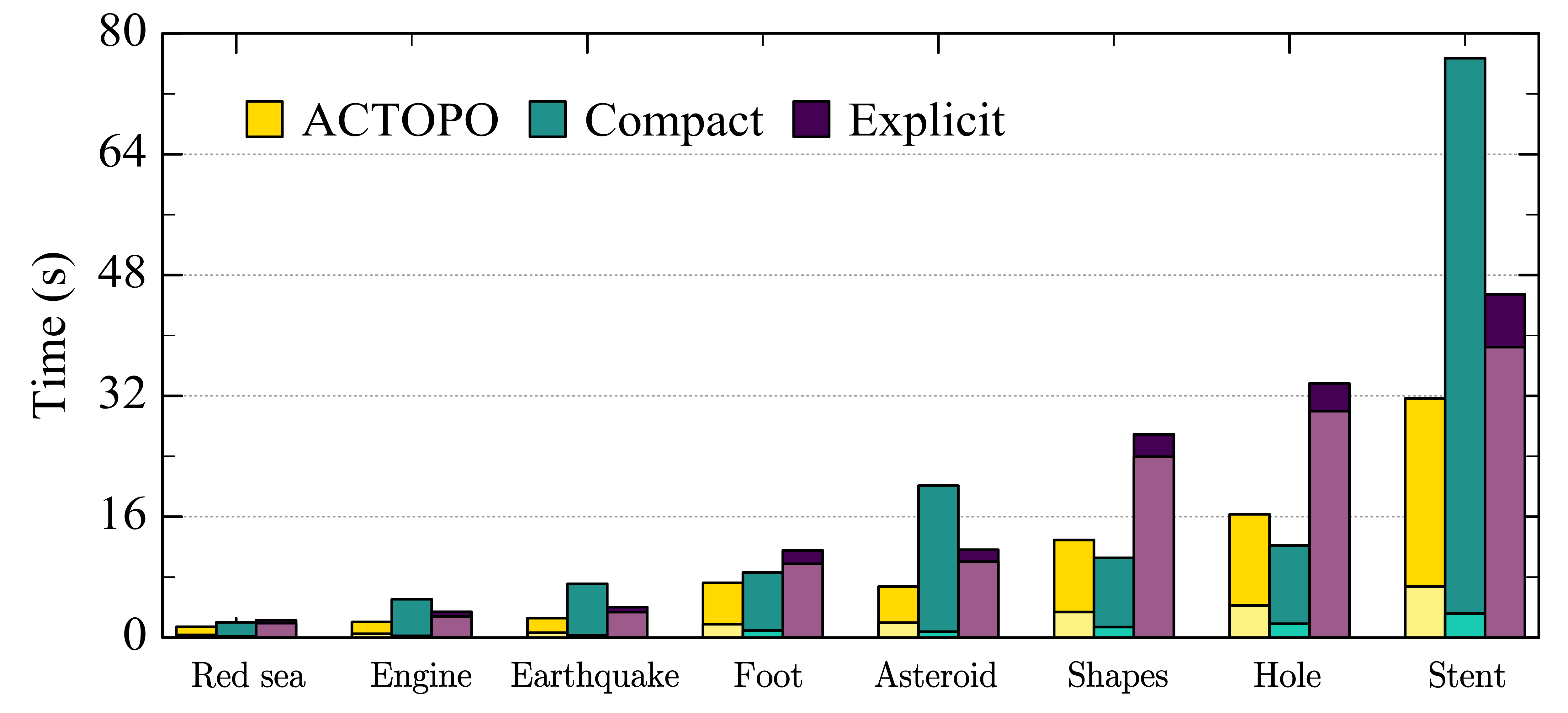} & 
        \includegraphics[width=\linewidth]{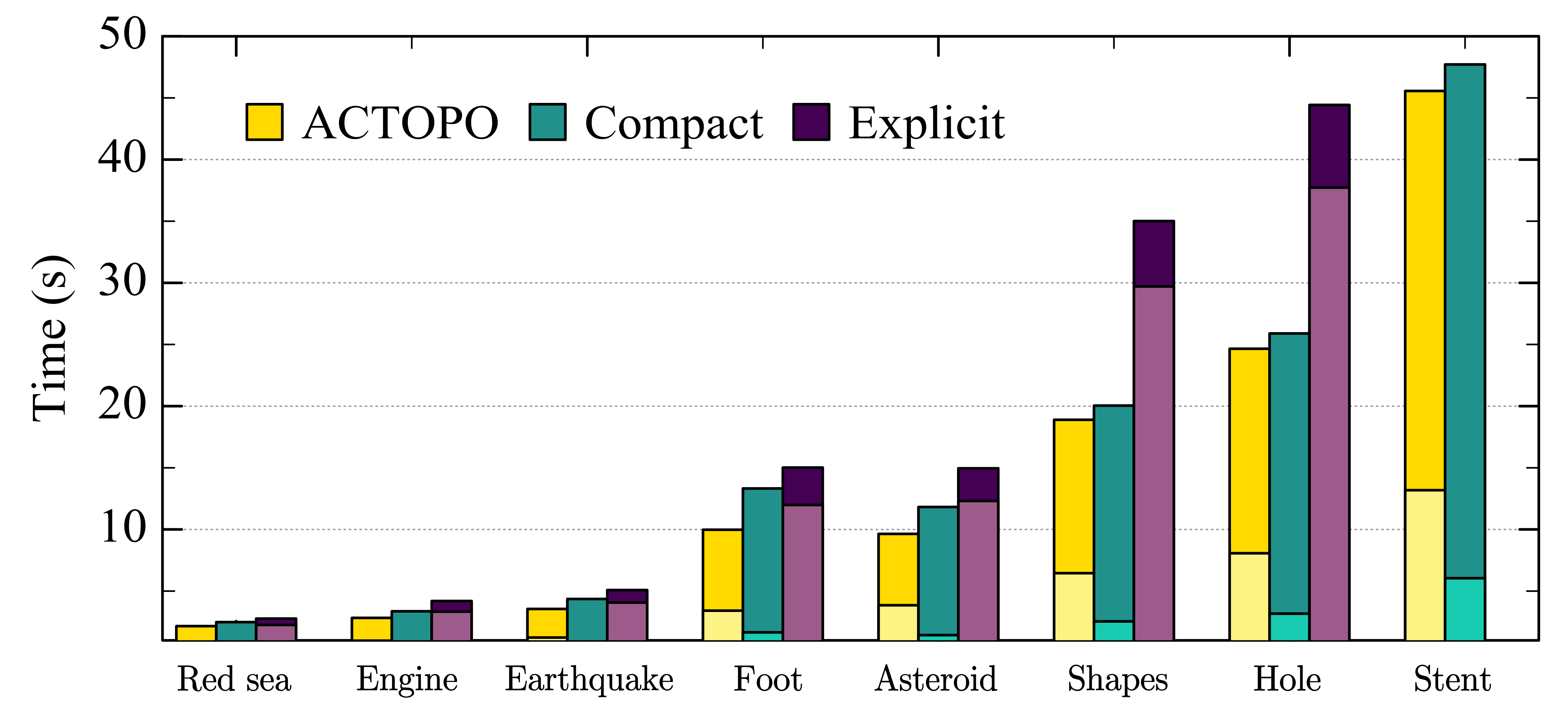} \\
        (a) & (b) & (c) \\
        \bottomrule
    \end{tabular}
    \caption{The total time (in seconds) used by three experimental plugins with different data structures: ACTOPO, TTK Compact Triangulation, and TTK Explicit Triangulation when the algorithm runs in parallel with OpenMP. The top bar indicates the time required by running the algorithm, while the bottom bar indicates the preprocessing time.}
    \label{fig:res_cmp_openmp}
\end{figure*}

\cref{fig:res_cmp_openmp} shows the total time and peak memory consumption. Notice that all algorithms are parallelized in a way that is agnostic to data structures (using OpenMP \cite{chandra2001parallel}). Specifically, for loops that iterate over simplices are split and distributed across threads. This is a challenge for block-based data structures like CompacTriangulation and ACTOPO. Suppose that the simplices indexed by a block are distributed between two threads, topological relations of that block will be computed and stored by each thread.

The resulting loss of performance is clearly visible in the results of \texttt{Test\-Topo\-Relations} (\cref{fig:res_cmp_openmp}(a)), where ACTOPO is about 1.2 times as fast as CompactTriangulation but 2.5 times as slow as ExplicitTriangulation.
However, this testing plugin already demonstrates the key advantage of block-based data structures, which is to speed up the preprocessing stage. During this step, ExplicitTriangulation is about 2.5 times as slow as ACTOPO. This advantage becomes fundamental depending on the topological relations required by the algorithm. For real plugins (see \cref{fig:res_cmp_openmp}(b) and (c)), Explicit Triangulation is always the slowest data structure, being 4.8 times as slow as ACTOPO at the preprocessing stage and requires 1.6 times to complete the execution on average. 

ACTOPO is generally the fastest data structure performing on par with CompactTriangulation. Exceptions are the Foot and Stent dataset, where CompactTriangulation performs sensibly worse than ACTOPO. This is because these two datasets contain many more critical points than other datasets, and CompactTriangulation is slower in computing boundary vertices than our data structure, as it needs to recompute triangle list for the same block multiple times while ACTOPO already has those information encoded at initialization time.

This is an impressive result if we consider that, with ACTOPO, the algorithm's instructions are executed by half the number of threads (6 instead of 12). This demonstrates the advantage that a task-parallel data structure can provide for unstructured mesh processing, leading to competitive time performance and efficient memory usage. 
ACTOPO constantly performs on par with CompactTriangulation using only 2\% more memory for \texttt{Scalar\-Field\-Critical\-Points} and \texttt{Test\-Topo\-Relations}, and 8\% less memory for \texttt{Discrete\-Gradient}. Notably, both data structures are at least 2 times as compact as ExplicitTriangulation. Given that using parallel algorithm has very limited effects on the peak memory usage, we provide detailed figures in supplemental materials.

\section{Conclusion}
\label{sec:conclusion}

In this paper, we present a new block-based task-parallel approach that integrates a producer-consumer paradigm to improve data structures' performance. A concrete implementation of this approach, called Accelerated Clustered Topological (ACTOPO) data structure, was also introduced and evaluated.
In our experimental evaluation, we test ACTOPO in two different scenarios. With sequential algorithms, ACTOPO provides a substantial speedup compared to state-of-the-art block-based data structures (TopoCluster \cite{Liu2021topocluster}). It provides timings comparable to static data structures (ExplicitTriangulation \cite{Tierny2018ttk}), using only half the memory. With parallel algorithms, ACTOPO maintains its compactness while providing the best time performance in most cases, when compared to other data structures.

\paragraph*{Limitations.}
The main limitation of ACTOPO is the spatial buffer. Our evaluation has shown that this type of buffer does not allow to completely eliminate the waiting time of the consumer thread when running the \texttt{Morse\-Smale\-Complex} plugin. Although the implementation of spatial buffer helps reduce the overall waiting time compared to the linear order, ACTOPO still faces difficulties in handling a processing algorithm exhibiting non-linear access patterns.

The main objective of ACTOPO is to provide an algorithm-agnostic interface for mesh processing. One possible direction for our future work is to explore learning-based methods to guide the producers in selecting the blocks to precompute. This may further reduce the waiting time for the consumer thread and result in faster algorithm execution. Another direction is to refine the block-based task-parallel model for specific algorithms. This will allow the development of new buffering strategies tailored to an algorithm's access patterns.


\section*{Supplemental Materials}
\label{sec:supplemental_materials}

All supplemental materials are available on OSF at \url{https://osf.io/b3d6k/}, released under a CC BY 4.0 license.
In particular, they include (1) Excel files containing the results obtained during experiments, (2) pseudocode for computing a representative set of localized topological relations with ACTOPO, and (3) figure images geenrated from the result data. 
The source code of ACTOPO data structure, including the plugins we used during the experiments, can be found at the same URL, released under BSD License.





\acknowledgments{
	The authors would like to thank Philips Research, Hamburg, Germany, for the Foot dataset, General Electric for the Engine dataset, and Michael Meißner, Viatronix Inc. for the Stent dataset. The Red sea dataset is courtesy of the Red Sea Modeling and Prediction Group (PI Prof. Ibrahim Hoteit). The Earthquake and Asteroid datasets are kindly shared by Mathieu Pont \cite{Pont2022Wasserstein}. The remaining tetrahedral meshes (Shapes, and Hole) are courtesy of Yixin Hu from New York University.
}

\bibliographystyle{abbrv-doi-hyperref}

\bibliography{references}


\appendix 







\end{document}